\documentclass[a4paper,fleqn,usenatbib]{mnras}
\usepackage[T1]{fontenc}
\usepackage{graphicx}
\usepackage{multirow}
\usepackage{txfonts}
\usepackage{color}
\usepackage{etoolbox}
\usepackage{hyperref}
\makeatletter

\title[Evolutionary modelling of sdB stars]{Evolutionary modelling of subdwarf B stars using MESA with the predictive mixing and convective premixing schemes}
\author[Ostrowski et al.]
{J. Ostrowski$^{1}$\thanks{E-mail: jakub.ostrowski@up.krakow.pl},
A. Baran$^{1,2,3}$, S. Sanjayan$^{1,4}$, S. K. Sahoo$^{1,4}$\\
$^{1}$ARDASTELLA Research Group, Institute of Physics, Pedagogical University of Krakow, ul. Podchor\k{a}\.zych 2, 30-084 Krak\'ow, Poland\\
$^{2}$Department of Physics, Astronomy, and Materials Science, Missouri State University, Springfield, MO 65897, USA\\
$^{3}$Embry-Riddle Aeronautical University, Department of Physical Science, Daytona Beach, FL\,32114, USA\\
$^{4}$Nicolaus Copernicus Astronomical Centre, Bartycka 18, 00-716 Warszawa, Poland}
\begin{document}

\date{Accepted ... Received ...; in original form ...}


\maketitle

\label{firstpage}

\begin{abstract}
  Results of evolutionary modelling of subdwarf B stars are presented. For the first time, we explore the core and near-core mixing in the subdwarf B stars using new algorithms available in the \texttt{MESA} code: the predictive mixing scheme and the convective premixing scheme. We show how both methods handle the problems with determination of convective boundary, discrepancy between core masses obtained from asteroseismology and evolutionary models, and long-standing problems related to the core-helium-burning phase such as splitting of the convective core and the occurrence of breathing pulses. We find that the convective premixing scheme is the preferable algorithm. The masses of the convective core in case of the predictive mixing and the combined convective and semiconvective regions in case of the convective premixing scheme are higher than in the models with only the Ledoux criterion, but they are still lower than the seismic-derived values. Both algorithms are promising and alternative methods of studying models of subdwarf B stars.
\end{abstract}


\section{Introduction} \label{sec_introduction}

Subdwarf B (sdB) stars are extreme horizontal branch stars undergoing helium burning in their cores. They are hot and compact with effective temperature $T_\mathrm{eff} = 20000 - 40000$ K, surface gravity $\log\,g = 5.0 - 5.8$ and radii $R = 0.15 - 0.35\,R_\odot$ \citep[][]{Heber09,Heber16}. The range of masses of hot subdwarfs is rather narrow, with $68.3\%$ of the sdB stars contained between $0.439$ and $0.501\,M_\odot$, with a median mass of $M = 0.471\,M_\odot$  \citep[the canonical mass,][]{Fontaine12}.

The characteristic feature of sdB stars is a very thin hydrogen envelope, $M_\mathrm{env} < 0.01\,M_\odot$ \citep{Heber86,Saffer94}, which is too thin to sustain a hydrogen-burning shell. Hence, contrary to the typical horizontal branch stars, they have only one energy source. After the helium is exhausted in the core, the sdB stars omit the evolution on the asymptotic giant branch and move directly to the white dwarf cooling track. Several evolutionary channels that lead to the removal of the envelope were proposed. More than half of the sdB stars are members of the short period binaries \citep{Maxted01,Copperwheat11} so the mass loss resulting from various interactions between components might be the dominant channel. There are various possibilities explaining the creation of single sdBs, such as merger events, hot-flash scenarios or the presence of sub-stellar companions. The evolutionary channels are discussed by, e.g., \citet{Mengel76,DCruz96,Han02,Han03,Miller08,Fontaine12,Charpinet18}.

Stellar oscillations are detected in many sdB stars. \citet{Charpinet96} predicted the existence of p-modes in sdBs and they were observationally discovered by \citet{Kilkenny97}. The g-modes in sdB stars were discovered later by \citet{Green03}. Typical periods of p-mode sdB pulsators are of the order of minutes and the periods of g-mode pulsators are of the order of hours \citep{Heber16,Holdsworth17,Reed18}. Both discoveries opened a way for applying asteroseismic techniques to probe the internal structure of sdBs.

Seismic studies of hot subdwarfs are usually performed using the forward modelling method \citep[e.g.,][]{Charpinet08,vanGrootel08} with static, structural models \citep[e.g.,][]{Brassard08,Brassard09}. Asteroseismology allowed determining the convective core masses of four g-mode pulsators: $M_\mathrm{cc} = 0.22 \pm 0.01\,M_\odot$ \citep[KPD 0629-0016;][]{vanGrootel10a}, $M_\mathrm{cc} = 0.28 \pm 0.01\,M_\odot$ \citep[KPD 1943+4058;][]{vanGrootel10b}, $M_\mathrm{cc} = 0.274^{+0.008}_{-0.010}\,M_\odot$ or $M_\mathrm{cc} = 0.225^{+0.011}_{-0.016}\,M_\odot$ \citep[KIC 02697388; two solutions obtained,][]{Charpinet11}, $M_\mathrm{cc} = 0.198 \pm 0.010\,M_\odot$ \citep[EC 21494-7018;][]{Charpinet19}. The first three stars have masses close to the canonical value, whereas the last star has significantly lower mass, $M_\mathrm{sdB} = 0.391 \pm 0.009\,M_\odot$, and the newer generation of static models were used in this case. All of the solutions point to rather young models with central helium abundance in the range of $Y_\mathrm{c} \approx 0.8 - 0.5$.

There is a problem of discrepancy between asteroseismic masses of convective cores and the masses obtainable from the evolutionary models. The seismic-derived core masses are much higher than the values yielded by the evolutionary models. With no additional mixing, the convective core has a constant mass of about $M_\mathrm{cc} \approx 0.1\,M_\odot$. The past efforts show that additional mixing, such as semiconvection or overshooting help to increase the size of the core \citep{Sweigart87,Dorman93}. Element diffusion also allows the core to grow \citep{Michaud07}. Other proposed solutions involve microphysics, such as changes to opacities or nuclear reaction rates. Comprehensive discussion on the subject and additional models can be found in \citet{Schindler15,Schindler17}. Some of these methods allow the core to grow up to $0.25\,M_\odot$, but for younger models (with high $Y_\mathrm{c}$), adequate for the mentioned seismic targets, the core masses are lower, $M_\mathrm{cc} < 0.2\,M_\odot$. The problem is still unresolved.

The core-helium-burning phase is a particularly challenging stage of stellar evolution. Due to the complex behaviour of the physical quantities (opacity, temperature, density, etc.), the radiative gradient, $\nabla_\mathrm{rad}$, develops a local minimum during this phase. This behaviour is well known and leads to problems with the models, such as splitting of the convective core \citep[e.g.,][]{Paczynski67,Castellani71,Eggleton72,Dorman93,Salaris17}. Another problem with core helium burning is the possibility of breathing pulses discussed in Appendix\,\ref{sec_breathing}. An excellent modern overview of the core-helium-burning phase can be found in \citet{Constantino15,Constantino16,Constantino17}.

The goal of this paper is to explore the behaviour of the core and near-core mixing using new algorithms available in the \texttt{MESA} code: the predictive mixing (PM) scheme and the convective premixing (CPM) scheme \citep{Paxton18,Paxton19}. They were not available when \citet{Ostensen12,Schindler15,Ghasemi17} and \citet{Xiong17} previously used \texttt{MESA} for modelling sdB stars, and \citet{Ratzloff19} and \citet{Kupfer20} used only the PM scheme. We aim to obtain models with an acceptable internal structure and core masses compatible with asteroseismic results, while eliminating or minimizing the problems related to core-helium burning.

The structure of this paper is as follows. In Section\,\ref{sec_sdB}, we explore evolutionary tracks for the selected models. In Sections\,\ref{sec_core} and \ref{sec_gradients}, we discuss properties of cores and analyse behaviour of gradients. Section\,\ref{sec_asteroseismology} contains comparison with asteroseismology. In Section\,\ref{sec_periods}, we compare period spacings of the models. Section\,\ref{sec_summary} contains conclusion. Appendix\,\ref{sec_convective_boundaries} presents problems with determining the convective boundaries and describes the PM and CPM algorithms. The physics of the calculated \texttt{MESA} models is presented in Appendix\,\ref{sec_models}. In Appendix\,\ref{sec_progenitors}, we show the basic properties of the progenitors.

\section{Models of sdB stars} \label{sec_sdB}

We calculated the evolutionary models using the \texttt{MESA} code \citep[Modules for Experiments in Stellar Astrophysics;][]{Paxton11,Paxton13,Paxton15,Paxton18,Paxton19}, version 11701. The CP and CPM algorithms are briefly described in Appendices\,\ref{sec_predictive} and \ref{sec_premixing}. The physical and numerical setup used in the models is presented in details in Appendix\,\ref{sec_models} for the sdB stars, and in Appendix\,\ref{sec_progenitors} for their progenitors.

The general changes to the evolutionary tracks caused by the changes of metallicity, mass of the helium core, mass of the envelope, etc. follow the same direction as in the previous study by \citet{Schindler15} that used the \texttt{MESA} code, version 7184. Inclusion of more advanced algorithms related to convective core boundaries do not change these basic properties of sdB models and hence we do not repeat this discussion here. Instead, we focus on the effects of the PM and the CPM schemes on the evolution of sdB stars and their cores.

\begin{figure}
  \includegraphics[clip,width=\columnwidth]{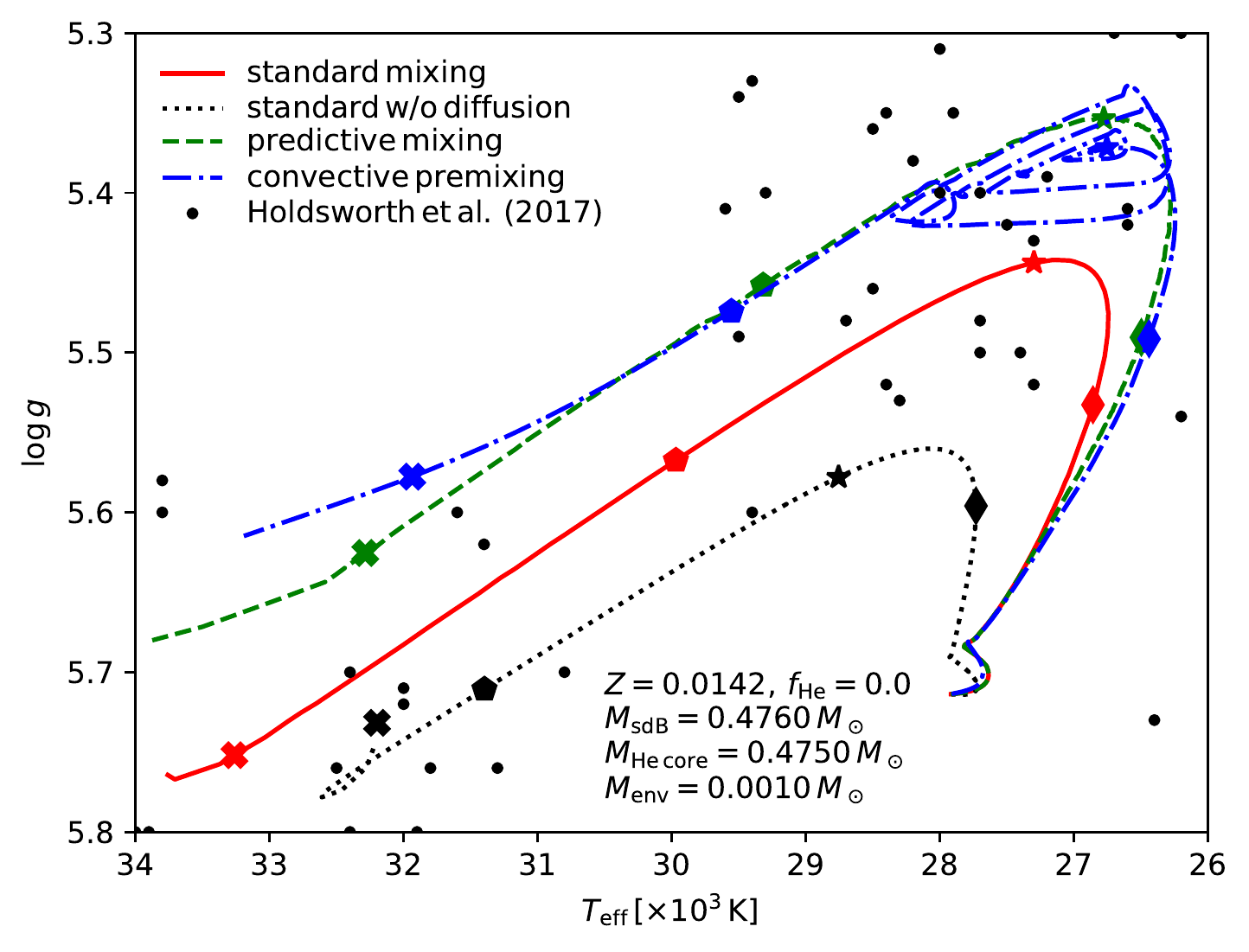}
  \caption{The $\log\,g$ vs. $\log\,T_\mathrm{eff}$ diagram with evolutionary tracks of sdB stars calculated with metallicity $Z=0.0142$, mass of helium core $M_\mathrm{He\,core} = 0.476M_\odot$ and envelope mass $M_\mathrm{env} = 0.001M_\odot$. The solid-red line depicts a model with standard mixing, the dotted-black line the model without diffusion, the dashed-green line the model with the PM scheme and the dashed-dotted-blue line the model with the CPM scheme. Diamonds, stars, pentagons and crosses on the tracks represent models with $Y_\mathrm{c}=0.5$, $0.1$, $0.01$ and $0.001$, respectively. The black dots represent pulsating subdwarfs from \citet{Holdsworth17}.}
  \label{fig_hr_comparison}
\end{figure}

In order to compare the effects of the considered mixing mechanisms we select three representative models for a detailed analysis. They are calculated from the progenitor with an initial mass of $M_\mathrm{i} = 1.0\,M_\odot$, the solar chemical composition, $Z = 0.0142$, $Y = 0.2703$, and mass of the helium core $M_\mathrm{He,\,core} = 0.475\,M_\odot$. The envelope mass is $M_\mathrm{env} = 0.001\,M_\odot$ and hence the total mass of the sdB is $M_\mathrm{sdB} = 0.476\,M_\odot$. Element diffusion is included and there is no overshooting from the convective core, $f_\mathrm{He} = 0.0$. The only difference between the models is the way how the boundary of the convective core is determined: the first one uses only the Ledoux criterion (standard mixing), the second utilizes the PM scheme and the third incorporates the CPM scheme. The results of this comparison are representative for the models with other metallicities and other masses of envelopes and helium cores.

The effects of including the PM and CPM schemes in the stellar calculations are shown in Figure\,\ref{fig_hr_comparison}, in which we compare evolutionary tracks in the $\log\,g$ vs. $\log\,T_\mathrm{eff}$ diagram for the models with standard mixing (the solid-red line), the PM scheme (the dashed-green line), the CPM scheme (the dashed-dotted-blue line) and with standard mixing, but with disabled diffusion (the dotted-black line). Diamonds, stars, pentagons and crosses on the tracks represent models with $Y_\mathrm{c}=0.5$, $0.1$, $0.01$ and $0.001$, respectively. The black dots represent pulsating sdBs from \citet{Holdsworth17}.

The tracks have typical shapes expected for sdB stars \citep[e.g.][]{Charpinet00}. They cover evolutionary stage from the full onset of the convective core to the point where $Y_\mathrm{c} < 1 \times 10^{-4}$ and the convective core vanishes.

The shapes of the evolutionary tracks are different for various mixing effects, which is a direct consequence of different sizes of the convective cores (Section\,\ref{sec_core}). While we do not focus here on models without atomic diffusion, it is worth showing how small is the track in such case when compared to the case with diffusion included. Without diffusion the convective core has small, constant size and hence this is a necessary process that should be included in all models \citep{Schindler15}.

Evolutionary tracks for models with the PM scheme or the CPM scheme have a few interesting properties. They are longer than the track with just the Ledoux criterion, which suggests that the additional mixing might indeed enhance the growth of the convective core. Their evolution in the H-R diagram is also very similar for most of their lifespan, up to the moment when central helium abundance is low, $Y_\mathrm{c} \approx 0.1$. Then, the track with the PM scheme proceeds smoothly, but a set of loops occurs for the track with the CPM scheme. This is a manifestation of the emergence of breathing pulses (Appendix\,\ref{sec_breathing}). From this point, we consider the further evolution of this model unrealistic, which is justified in the following sections.

\section{Convective cores} \label{sec_core}

\begin{figure*}
  \includegraphics[clip,width=\textwidth]{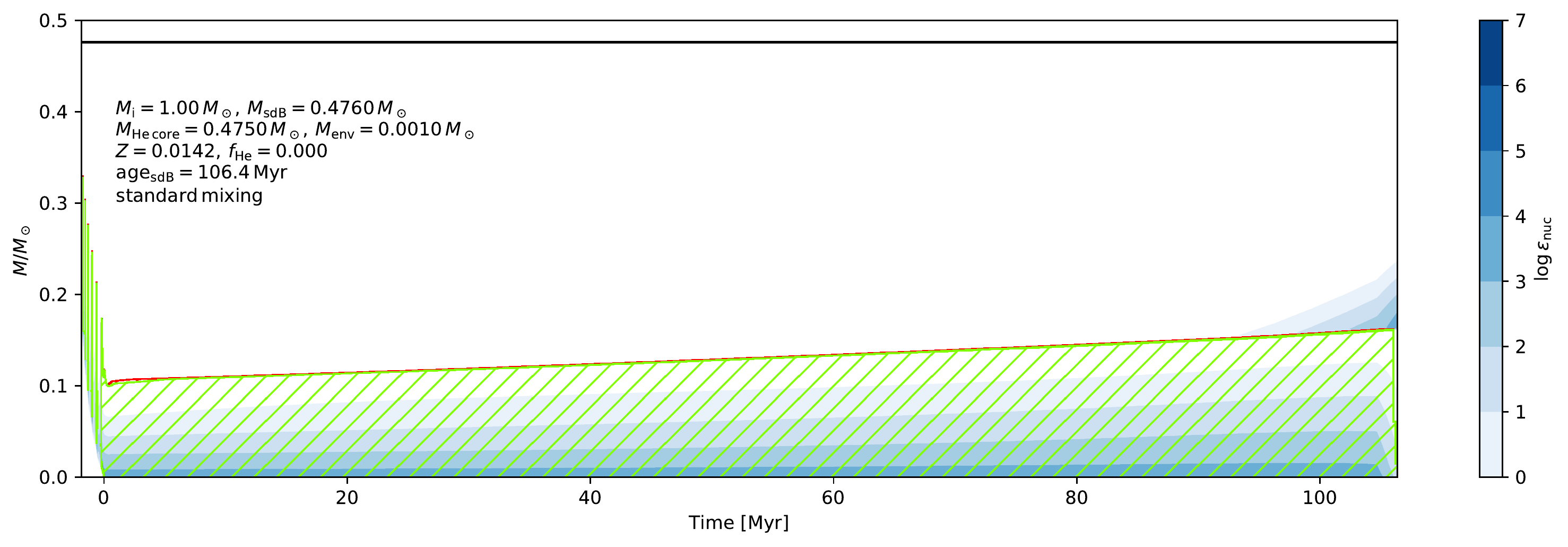}
  \includegraphics[clip,width=\textwidth]{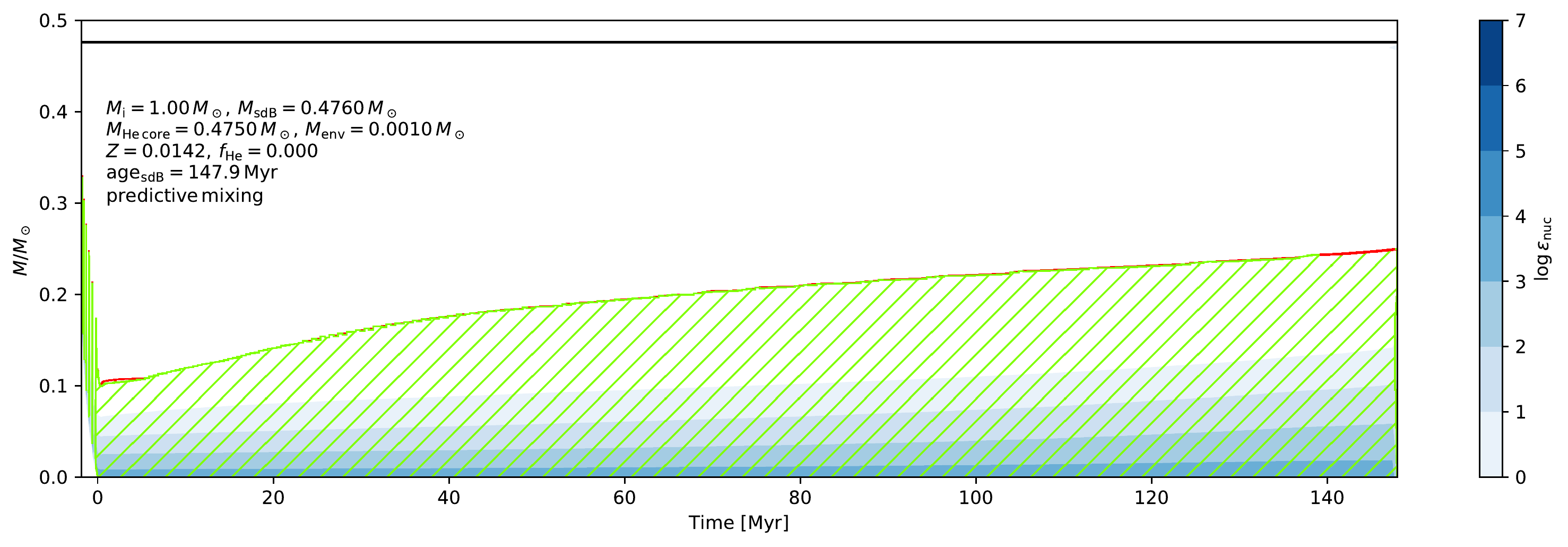}
  \includegraphics[clip,width=\textwidth]{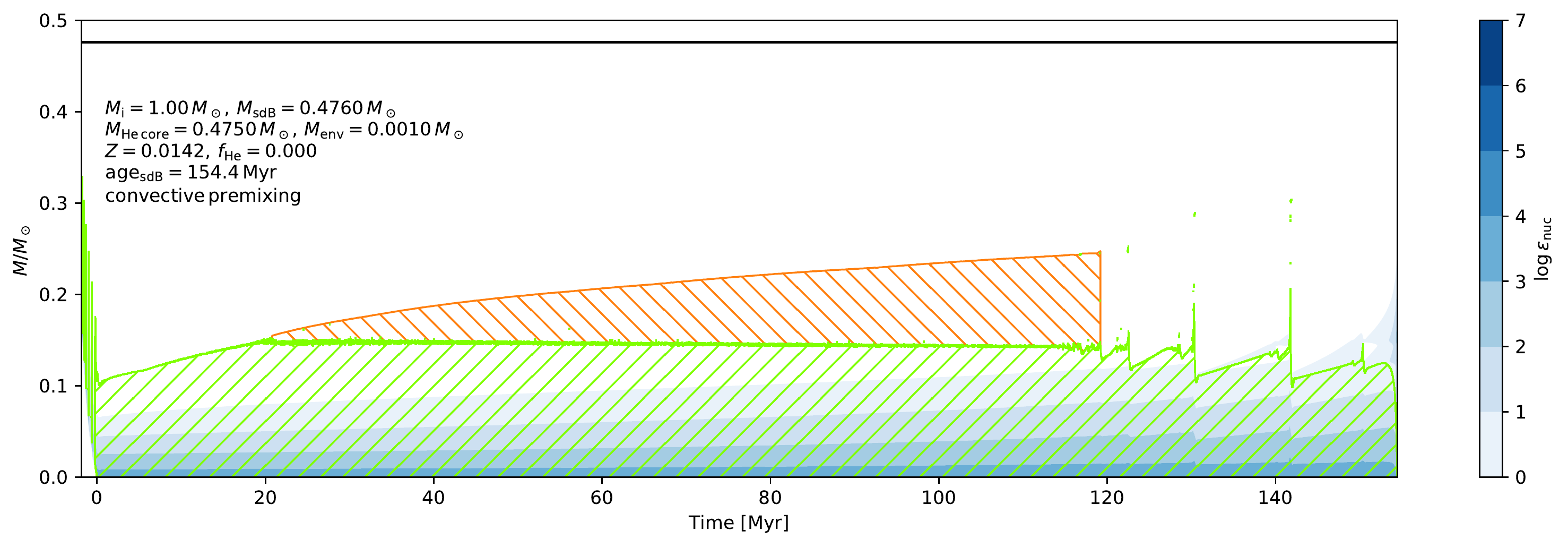}
  \caption{Kippenhahn diagrams for models with the standard mixing (top panel), the PM (middle panel) and the CPM schemes (bottom panel), calculated with metallicity $Z=0.0142$, mass of helium core $M_\mathrm{He\,core} = 0.476\,M_\odot$ and envelope mass $M_\mathrm{env} = 0.001\,M_\odot$. The green hatched lines depict convective zones, the red hatched lines show semiconvective zones in the sense of \citet{Kato66} (not in the bottom panel, details in text), and the orange hatched lines show the semiconvective zone by definition of \citet{Schwarzschild58}. The solid black line shows the surface of a star. The rate of nuclear energy generation, $\log\,\epsilon_\mathrm{nuc}$, is shown in shades of blue. The structure of the star is shown in a function of time elapsed since the start of the sdB phase.}
  \label{fig_kipp}
\end{figure*}


The differences between tracks presented in Figure\,\ref{fig_hr_comparison} can be explained by the different sizes and evolution of the convective cores caused by algorithms applied to search for convective boundaries. Changes of the internal structure of a star during its evolution can be followed using the Kippenhahn diagrams. In Figure\,\ref{fig_kipp} we show such diagrams for the same three representative models that are shown in Figure\,\ref{fig_hr_comparison}: the standard model in the top panel, the model with the PM scheme in the middle panel and the model with the CPM scheme in the bottom panel. The model without diffusion is not shown because it is too simple and not relevant for further analysis. The green hatched lines depict convective zones, the red hatched lines show semiconvective zones in the sense of \citet{Kato66}, which is related to the Ledoux criterion, and the orange hatched lines show the semiconvective zone in the sense of \citet{Schwarzschild58}, which can occur if the CPM scheme is used (cf. Appendix\,\ref{sec_premixing}). The rate of nuclear energy generation, $\log\,\epsilon_\mathrm{nuc}$, is shown in shades of blue. The structure of a star is shown in a function of time elapsed since the start of the sdB phase, until the depletion of helium  and disappearance of convection in the core. Note the different ranges in the abscissae in the panels. Plotting of semiconvective regions in the sense of \citet{Kato66} and \citet{Langer83} is omitted in models with the CPM scheme. We found that \texttt{MESA} reports tiny semiconvective regions that are the result of small noise in the radiative gradient, $\nabla_\mathrm{rad}$. They are not significant for the model, but they would suppress the legibility of the plot.

Comparison between the standard model (the top panel of Figure\,\ref{fig_kipp}) and the model utilizing the PM scheme (the middle panel) shows significant difference in the growth of their convective cores. The initial mass of the convective core is $M_\mathrm{cc} \approx 0.106\,M_\odot$ in all three cases. In the case of standard mixing a slight growth of the core to the maximum value of $M_\mathrm{cc} \approx 0.161\,M_\odot$ can be seen at the end of the core-helium-burning phase. The value is below the asteroseismic predictions (Section\,\ref{sec_introduction}) and clearly shows that there is a problem with too small convective cores in the standard evolutionary calculations of sdB stars. On the other hand, the PM scheme allows the core to grow faster and to higher masses. For instance, for the age of $50\,\mathrm{Myr}$, the masses of convective cores are $M_\mathrm{cc} = 0.127\,M_\odot$ and $M_\mathrm{cc} = 0.186\,M_\odot$ for the cases with only the Ledoux criterion and with the PM scheme, respectively. The total core-helium-burning phase is also significantly longer with the PM scheme ($147.9\,\mathrm{Myr}$) than for the case with the standard mixing ($106.4\,\mathrm{Myr}$). In the standard model, $\log\,\epsilon_\mathrm{nuc}$ increases outside the convective core near the end of the core-helium-burning phase. This is the onset of the helium burning in the shell.

The structure of the presented models with standard mixing and the PM scheme is rather simple during the whole sdB stage, with a fully mixed convective core, very thin semiconvective layer (related to the Ledoux criterion) and the transition to the radiative envelope. The situation changes in the model with the CPM scheme (the bottom panel of Figure\,\ref{fig_kipp}). For the first $20$ million years, until the central helium abundance drops to $Y_\mathrm{c} \approx 0.75$, the evolution proceeds similarly to the case with the PM scheme. Then, a semiconvective zone \citep[in the sense of][]{Schwarzschild58} emerges and it grows with the evolution of a star. There is no additional prescription for semiconvective mixing in this region, as in e.g. models of \citet{Sweigart87} or \citet{Constantino15}, but the mixing is a consequence of the CPM scheme. Growth of the convective core ceases at the mass $M_\mathrm{cc} \approx 0.147\,M_\odot$ and its mass very slightly decreases to the value of $M_\mathrm{cc} \approx 0.143\,M_\odot$ at the age of $118.2$ million years. For this age, the mass at the top of semiconvective zone is $M_\mathrm{sc} \approx 0.249\,M_\odot$.

The semiconvective zone is plotted until the model achieves the age of $118.2$ million years, which corresponds to the central helium abundance $Y_\mathrm{c} = 0.098$. At this point the first breathing pulse occurs and we consider the following evolution as not realistic. Firstly, we consider the breathing pulses to be numerical artefacts, and secondly, the semiconvective zone becomes ill-defined. The way the CPM scheme works leads to a gradually increasing partially-mixed zone. When a breathing pulse occurs, there is a sudden, short-term increase of the size of the convective core, which is visible in the bottom panel of Figure\,\ref{fig_kipp}. If a part of the semiconvective zone becomes convective, it is instantly mixed with the rest of the core and the smooth, gradually built structure is lost after the breathing pulse ends. Hence, we simply do not consider semiconvective zone after the first pulse. A similar approach was used for the standard helium-burning objects by \citet{Paxton19}.

\begin{figure}
  \includegraphics[clip,width=\columnwidth]{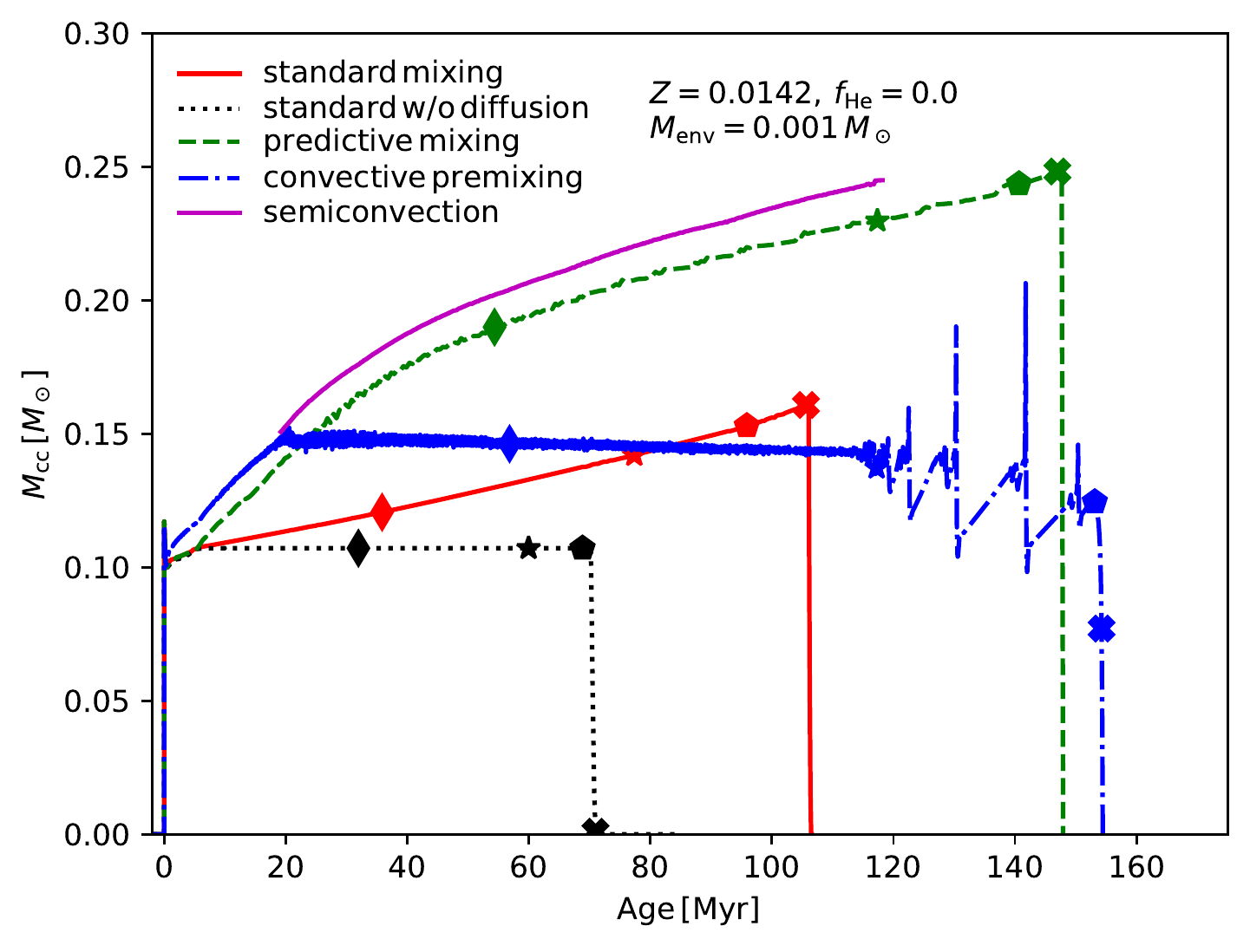}
  \caption{Masses of convective cores, $M_\mathrm{{cc}}$, in the function of age for the same models as in Figure\,\ref{fig_hr_comparison}. The solid-red line depicts the model with the standard mixing, the dotted-black line the model without diffusion, the dashed-green line the model with the PM scheme and the dashed-dotted-blue line the model with the CPM scheme. The solid-magenta line shows the top boundary of the semiconvective zone for the models with the CPM scheme. The models with $Y_\mathrm{c}=0.5$, $0.1$, $0.01$ and $0.001$ are marked by diamonds, stars, pentagons and crosses, respectively.}
  \label{fig_mass_cc}
\end{figure}

Changes of the masses of convective cores of the discussed models can be easily compared in Figure\,\ref{fig_mass_cc}, in which we present core masses, $M_\mathrm{{cc}}$, in function of age for the same models as in Figure\,\ref{fig_hr_comparison}. The solid-red line depicts the model with the standard mixing, the dotted-black line the model without diffusion, the dashed-green line the model with the PM scheme and the dashed-dotted-blue line the model with the CPM scheme. Additionally, the solid-magenta line shows the top boundary of the semiconvective zone for the model with the CPM scheme. The differences in total age of the sdB phases between different models can be immediately seen. The most basic model with only Ledoux criterion and no diffusion has a lifespan of just $71$ million years, whereas the standard mixing model with diffusion ends the sdB evolution after $106.4$ million years. The PM scheme significantly prolongs the lifespan to $147.9$ million years. In the case of the CPM scheme the total length of the sdB phase is even longer, $154.4$ million years, but this is mainly due to the breathing pulses that ingest fresh helium into the core. The more significant value is the age of $118.2$ million years corresponding to the first breathing pulse and hence the last useful model of the evolutionary track. At this age the abundance of helium in the core is $Y_\mathrm{c} = 0.098$. The model with the PM scheme achieves this value of $Y_\mathrm{c}$ at the age of $117.9$ million years, so the time of the evolution is in fact very similar for those two algorithms. All the models that we consider in this paper are calculated with the Ledoux criterion for convection, but results for the PM and CPM models calculated with the Schwarzschild criterion should be very similar \citep{Paxton18,Paxton19}.

The evolution of the top of semiconvective region, $M_\mathrm{{sc}}$, in the case of the CPM scheme follows rather closely the evolution of the mass of convective core for the PM model, which is shown in both Figure\,\ref{fig_kipp} and Figure\,\ref{fig_mass_cc}. For the part of the evolution when the value is well defined, $M_\mathrm{{sc}}$ is higher than $M_\mathrm{{cc}}$ by about $0.009 - 0.014\,M_\odot$, at the same age. The fully or partially mixed region consisting of a smaller convective core and a semiconvective zone on top of it seem to have a similar effect on properties of the star as a fully-mixed large convective core of a similar size. We can see that the age and evolutionary tracks (Figure\,\ref{fig_hr_comparison}) do not differ significantly between the PM and the CPM schemes, despite the important difference in the internal structure.

\begin{figure}
  \includegraphics[clip,width=\columnwidth]{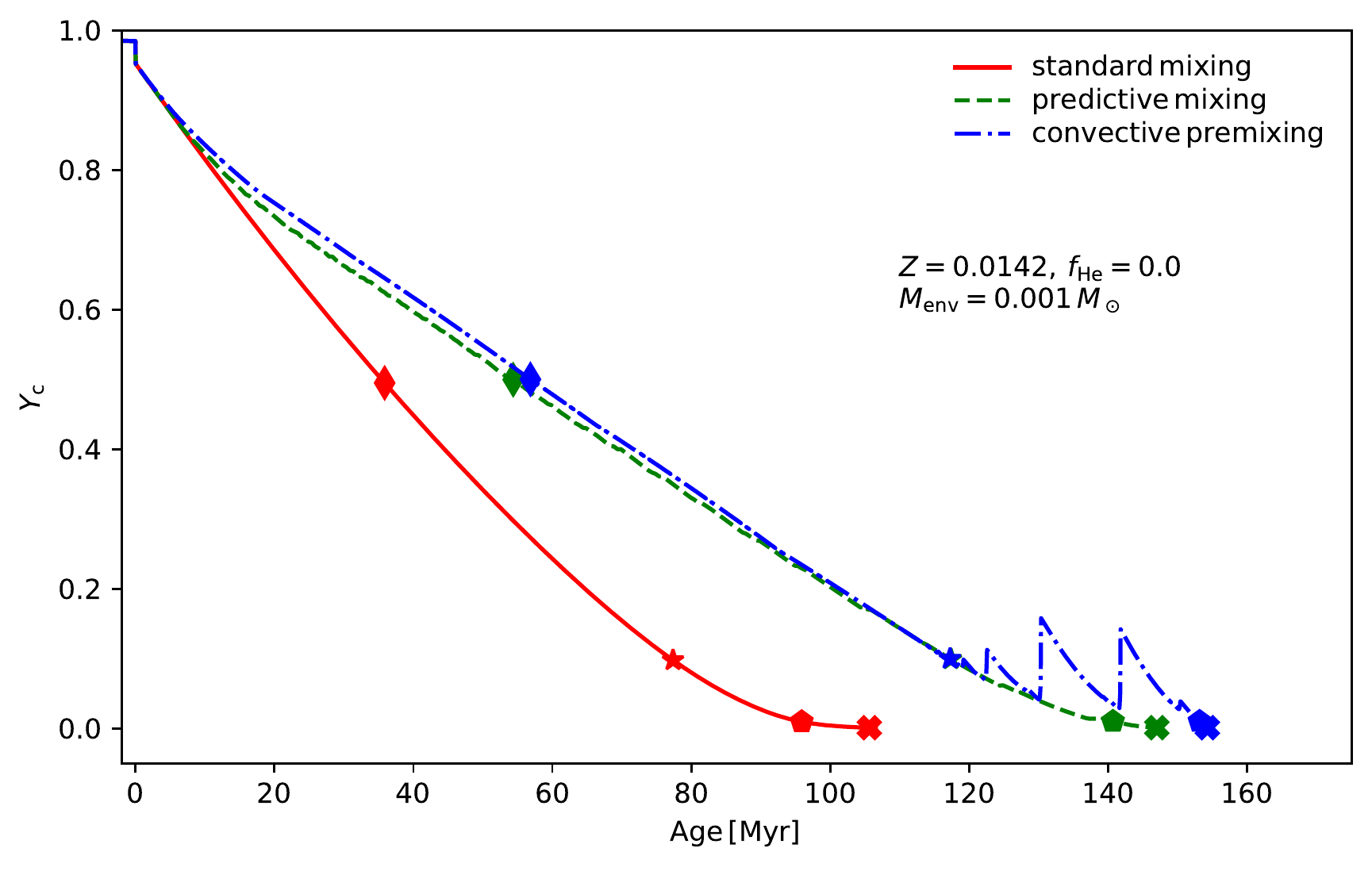}
  \caption{Central abundance of $^4$He, $Y_\mathrm{c}$, in function of age for the models with standard mixing (solid-red line), the PM scheme (dashed-green line) and the CPM scheme (dashed-dotted-blue line). Diamonds, stars, pentagons and crosses on the tracks represent models with $Y_\mathrm{c}=0.5$, $0.1$, $0.01$ and $0.001$, respectively.}
  \label{fig_center_he4}
\end{figure}

This conclusion is applicable only up to the occurrence of the breathing pulses, which are well visible in Figure\,\ref{fig_mass_cc}. The largest breathing pulse increases the mass of the convective core from $0.140\,M_\odot$ to $0.206\,M_\odot$ within just $130$ thousand years. The underlying cause of the increase of mass of a core is ingestion of additional helium into the core (Appendix\,\ref{sec_breathing}), which is illustrated in Figure\,\ref{fig_center_he4}, where $Y_\mathrm{c}$ is plotted in function of age for the same models as in Figure\,\ref{fig_kipp}: with standard mixing (solid-red line), the PM scheme (dashed-green line) and the CPM scheme (dashed-dotted-blue line). The value of $Y_\mathrm{c}$ should decrease monotonically with evolution during the core-helium-burning phase. For the considered models, this is the case for the standard mixing and the PM scheme. The CPM model behaves in such way to the point when $Y_\mathrm{c} = 0.098$, then five steep increases of $Y_\mathrm{c}$ are visible. During the phase of breathing pulses, the central helium abundance increases significantly, to $Y_\mathrm{c} = 0.158$ after the third pulse.

The evolution of the border of the convective core in the CPM model shows a slight numerical noise, despite very high spatial and temporal resolution required by the CPM scheme (Appendix\,\ref{sec_models}). This is similar to the behaviour of the core-helium-burning CPM model discussed by \citet{Paxton19}. In Figure\,\ref{fig_mass_cc}, it is shown that enabling diffusion in the model with standard mixing allows slow growth of the convective core and prolongs the lifespan of the model, when compared to the model without diffusion. This is in agreement with the results of \citet{Michaud07} and \citet{Schindler15}. Nevertheless, the standard model is not fully resolved temporally. The element-diffusion mixing, driven by the composition gradient at the core boundary, causes the convective core to grow, but in order to resolve this process properly in time, the time steps of the order of the convective mixing timescale would be necessary. They would be too short for stellar evolution calculations and hence it is virtually impossible to converge a model with standard mixing to a reasonable solution. This is another reason why the PM scheme was introduced to \texttt{MESA} \citep{Paxton18}.

\subsection{Problems with overshooting from helium core} \label{sec_overshooting}

\begin{figure*}
  \includegraphics[clip,width=\textwidth]{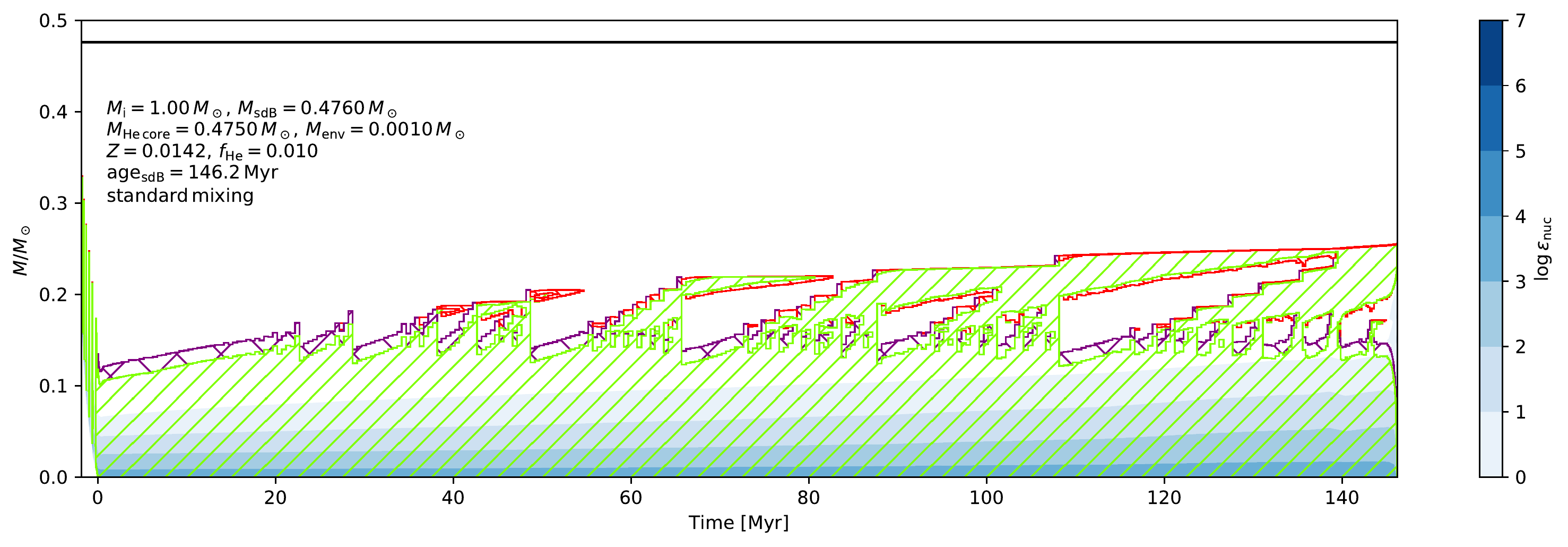}
  \includegraphics[clip,width=\textwidth]{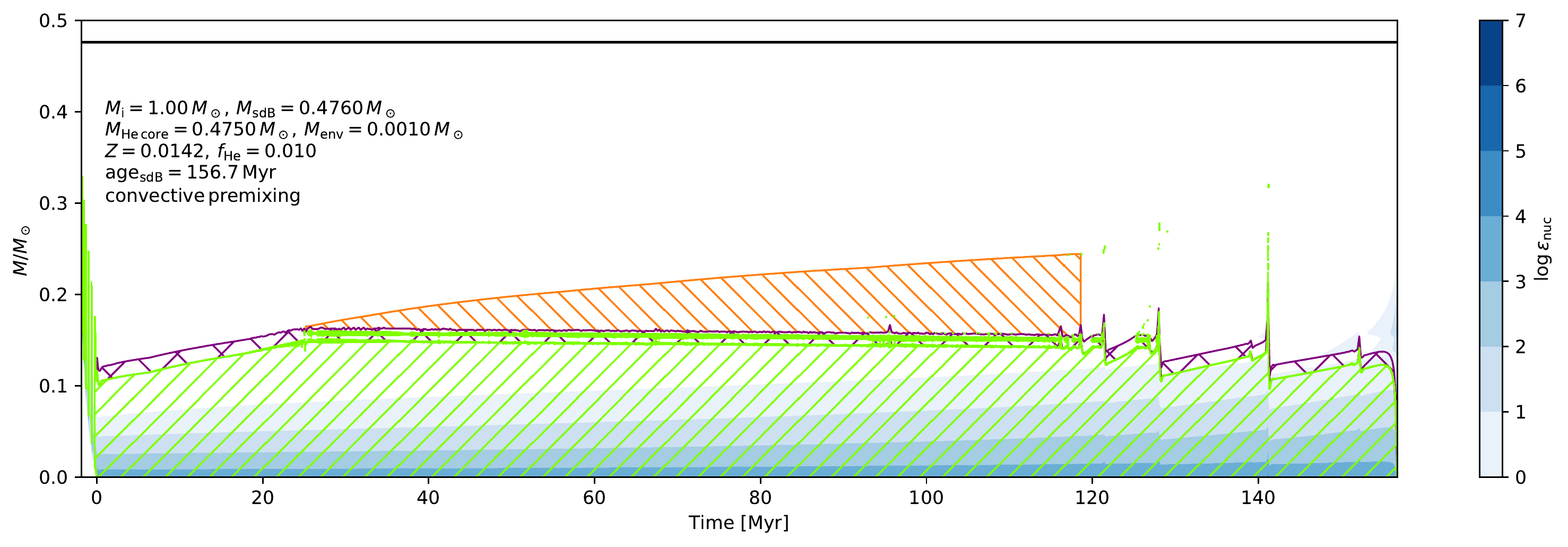}
  \caption{Kippenhahn diagrams for models with convective overshooting from the helium core, $f_\mathrm{He} = 0.01$ and with only the Ledoux criterion (top panel) and the CPM scheme (bottom panel). Other parameters and the meaning of symbols and colours are the same as for the models in Figure\,\ref{fig_kipp}. The violet cross-hatched lines show the regions with overshooting.}
  \label{fig_kipp_overshooting}
\end{figure*}

Including of overshooting from the helium-burning core leads to significant numerical issues in the \texttt{MESA} code. In Figure\,\ref{fig_kipp_overshooting}, Kippenhahn diagrams for models with convective overshooting from the helium core with efficiency $f_\mathrm{He} = 0.01$ and the other parameters are the same as for the models shown in Figure\,\ref{fig_kipp}. The meaning of symbols and colours is the same as in Figure\,\ref{fig_kipp} and the overshooting regions are shown with the violet cross-hatched lines.

\begin{figure}
  \includegraphics[clip,width=\columnwidth]{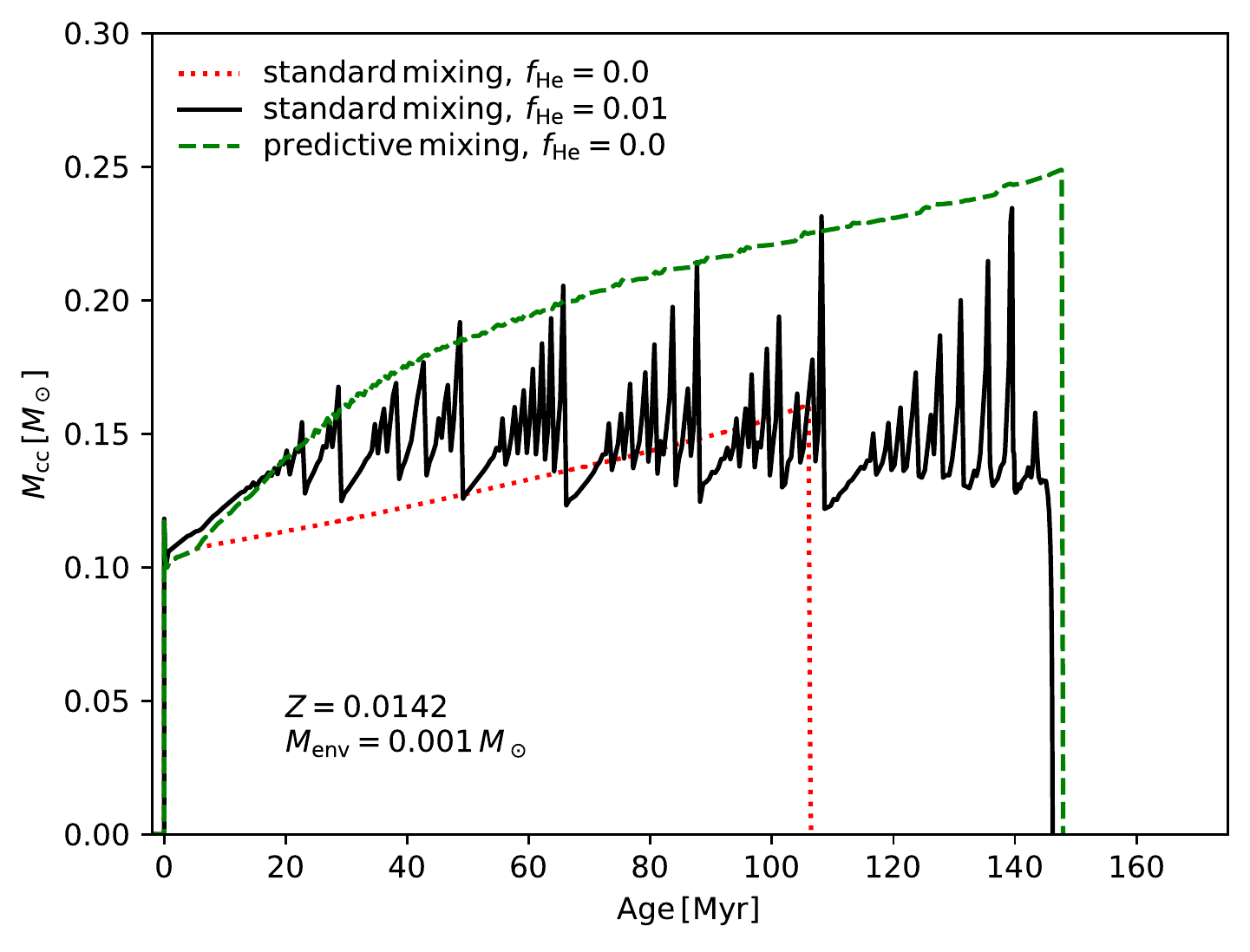}
  \caption{Masses of convective cores, $M_\mathrm{cc}$, in function of age for the model with the Ledoux criterion and overshooting efficiency from the helium core $f_\mathrm{He} = 0.01$ (solid-black line; the same model as in the top panel of Figure\,\ref{fig_kipp_overshooting}) and two reference models without overshooting: with the standard mixing (dotted-red line) and with the PM scheme (dashed-green line).}
  \label{fig_mass_cc_overshooting}
\end{figure}

The model with the Ledoux criterion is shown in the top panel of Figure\,\ref{fig_kipp_overshooting}. Kippenhahn diagram for the model with the PM scheme and overshooting is not shown because the structure of the star is almost the same as in the case with the Ledoux criterion only. The structure of the model is significantly different versus the case with $f_\mathrm{He} = 0.0$. The boundary of the convective core is very unstable, which is better illustrated in Figure\,\ref{fig_mass_cc_overshooting}, in which we show the mass of the convective core, $M_\mathrm{cc}$, in function of age, for the considered model with overshooting efficiency $f_\mathrm{He} = 0.01$ and, for reference, for previously considered models without overshooting and with and without the PM scheme. Significant numerical noise was also visible in the \texttt{MESA} models with overshooting considered by \citet{Schindler15}. In the Kippenhahn diagram, we can see that not only is the boundary of the convective core unstable, but also when the core is growing it often splits. That leads to a configuration with convective core and convective shell. We can see in the top panel of Figure\,\ref{fig_kipp_overshooting} and in Figure\,\ref{fig_mass_cc_overshooting} that the outline of the appearing convective shells follows the growth of the monolithic convective core or the semiconvective region in the case of the PM and the CPM schemes, respectively, with no overshooting applied.

In the case of the model with the CPM scheme, shown in the bottom panel of Figure\,\ref{fig_kipp_overshooting}, the inclusion of overshooting yields a structure similar to the case with $f_\mathrm{He} = 0.0$ (the bottom panel of Figure\,\ref{fig_kipp}). The core has a nearly constant size and an extended semiconvective zone is present. The biggest difference is the presence of the overshooting region on top of the convective core. The drawbacks of overshooting are less significant than in the cases of pure Ledoux criterion or the PM scheme, but they are still visible. There is a very thin convective zone inside of the overshooting region and there are a few short-term increases of the mass of the convective core before the breathing pulses. The total lifespan of the model is $156.7$ million years, which is $2.3$ million years longer than without overshooting. The time when the first breathing pulse occurs is $118.7$ million years, which is very similar to $118.2$ million years in the case with $f_\mathrm{He} = 0.0$. Most importantly, the mass at the top of the semiconvective zone is exactly the same regardless of overshooting or not. Therefore, overshooting does not provide any additional expansion of the mixed region when the CPM scheme is used.

The considered efficiency of overshooting, $f_\mathrm{He} = 0.01$, would be a rather small value during the main-sequence evolution \citep[e.g.,][]{Claret17}, but there are no calibrations performed for the core-helium-burning phase. The presented results are representative and we obtain similar structure for $f_\mathrm{He} = 0.005$ and $0.02$. One of the goals of this paper is to obtain models with acceptable structure and without significant numerical artefacts. Because of the issues with an unstable boundary of the convective core and irregular convective zones in the models with overshooting and pure Ledoux criterion, we reject models with such structure and do not consider them for further analysis and use. In the case of models with the CPM scheme the problems with overshooting are less pronounced, but they exist nevertheless, and it seems that including overshooting is not beneficial. For these reasons and the fact that the calibrations of overshooting for sdB or horizontal-branch stars do not exist, we opt for not using overshooting from the convective core in any of our models. \citet{Ghasemi17} calculated \texttt{MESA} models of sdBs with core overshooting and a structure similar to the structure of our overshooting models. \citet{Ghasemi17} do not treat the obtained structure as dubious and use the models for pulsation calculations. The additional acoustic cavities, which occur with the emerging convective shells, are potentially attractive for asteroseismology, as they lead to strong mode trapping. Nevertheless, here we consider such models as faulty and prefer to find other solutions that yield trapped modes, e.g., a non-homogenous semiconvective zone in models with the CPM scheme (cf. Section\,\ref{sec_periods}).

\section{Behaviour of gradients} \label{sec_gradients}

\begin{figure}
  \includegraphics[clip,width=\columnwidth]{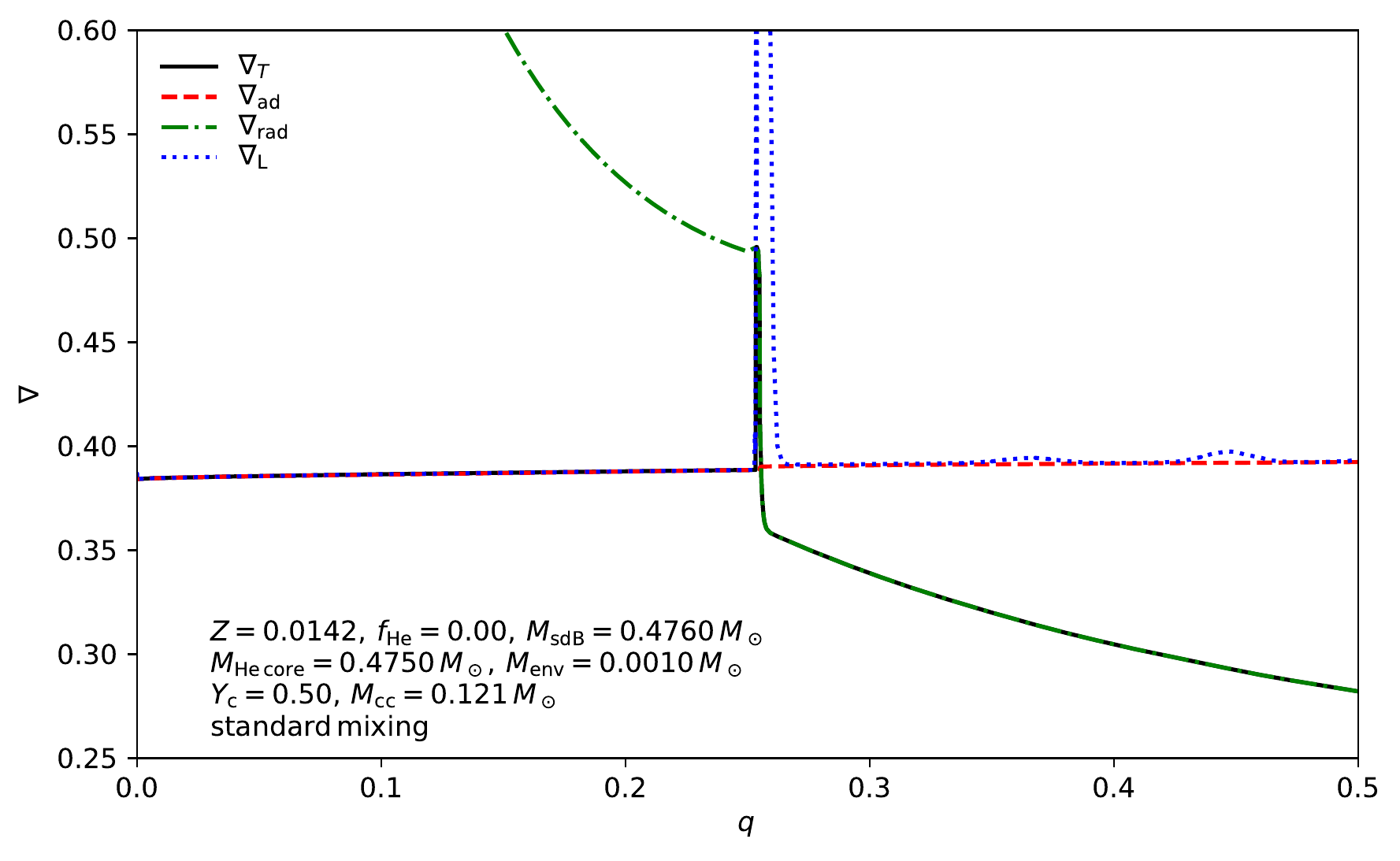}
  \includegraphics[clip,width=\columnwidth]{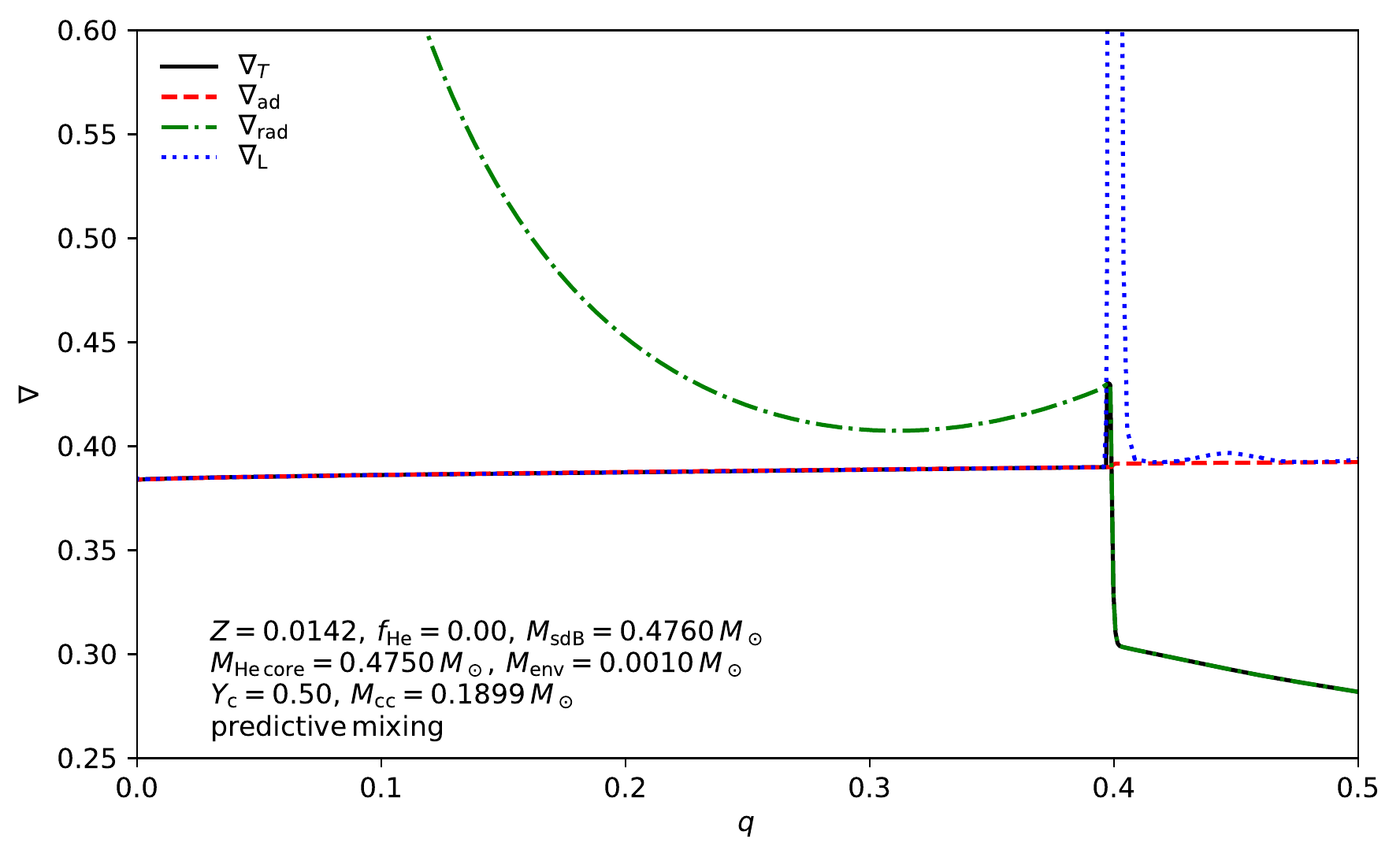}
  \includegraphics[clip,width=\columnwidth]{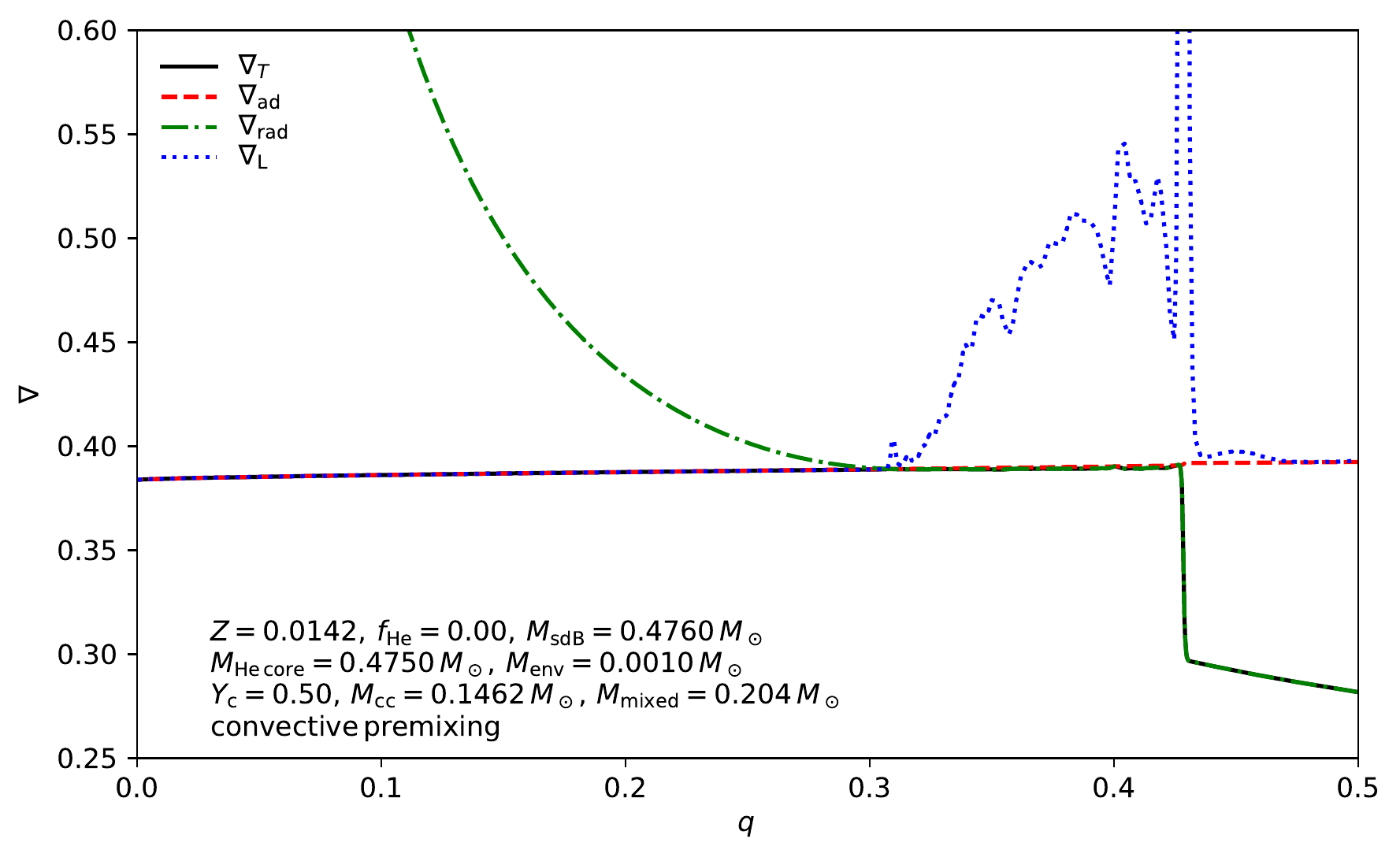}
  \includegraphics[clip,width=\columnwidth]{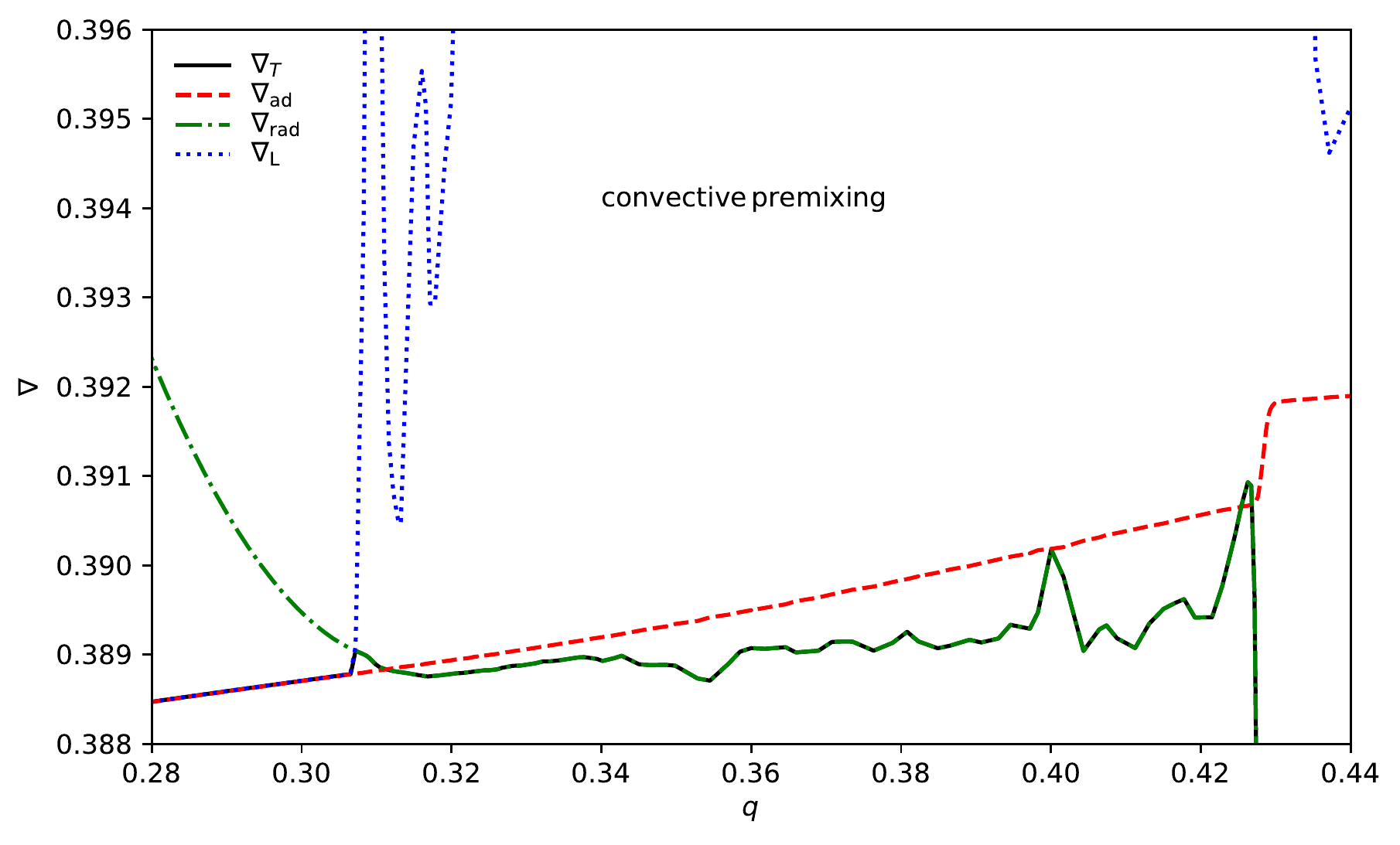}
  \caption{Comparison of actual temperature gradient $\nabla_\mathrm{T}$ (solid-black line), adiabatic gradient $\nabla_\mathrm{ad}$ (dashed-red line), radiative gradient $\nabla_\mathrm{rad}$ (dashed-dotted-green line), and Ledoux gradient $\nabla_\mathrm{L}$ (dotted-blue line), in function of relative mass, $q$, for selected sdB models with central helium abundance $Y_\mathrm{c} = 0.5$. Top panel: model with the standard mixing, second panel from top: model with the PM scheme, third panel from top: model with the CPM scheme, bottom panel: model with the CPM scheme, but with close-up of the semiconvective region.}
  \label{fig_gradients}
\end{figure}

Analysing gradients inside the models allows for better understanding of internal structure and differences discussed in the previous sections. In Figure\,\ref{fig_gradients} we compare the actual temperature gradient $\nabla_\mathrm{T}$ (solid-black line), adiabatic gradient $\nabla_\mathrm{ad}$ (dashed-red line), radiative gradient $\nabla_\mathrm{rad}$ (dashed-dotted-green line), and Ledoux gradient $\nabla_\mathrm{L}$ (dotted-blue line), in function of relative mass, $q = m/M$, where $m$ is the mass within a shell and $M$ is the total mass of a star. The selected sdB models have central helium abundance $Y_\mathrm{c} = 0.5$ and are chosen from the models presented in Figures\,\ref{fig_hr_comparison} - \ref{fig_center_he4}. The results for models with the standard mixing, the PM scheme, and the CPM scheme are shown in the top, second and third from top panels, respectively. The plots are limited to $q \leq 0.5$ and $\nabla \in [0.25, 0.60]$, to put focus on the features of the core and the semiconvective zone. In the bottom panel, we show the close-up of the semiconvective region for a model with the CPM scheme.

In every considered case, there is a very steep local maximum of the Ledoux gradient at the border of the convective core or, in the case of the CPM scheme, the semiconvective region. It is directly related to the sudden change of the abundance gradient caused by transition from a fully- or partially-mixed zone to a radiative zone with a uniform chemical composition. The small local maxima of $\nabla_\mathrm{L}$, visible on the radiative side, are the remnants of convective zones during helium subflashes (cf. Appendix\,\ref{sec_progenitors_flash}).

\citet{Gabriel14} argues that consistency with the mixing length theory (MLT) framework requires gradients neutrality on the convective side (Appendix\,\ref{sec_convective_boundaries}). The \texttt{MESA} models with the standard mixing have problems with fulfilling this requirement, as can be seen in the top panel of Figure\,\ref{fig_gradients}. At $q = 0.253$ there is a sharp transition between the convective core, where $\nabla_\mathrm{rad} > \nabla_\mathrm{L}$, and the sub-adiabatic radiative region, with a very thin ($\mathrm{d}q \approx 0.002$) semiconvective layer, where $\nabla_\mathrm{L} > \nabla_\mathrm{rad} > \nabla_\mathrm{ad}$. The difference between gradients, $\nabla_\mathrm{rad} - \nabla_\mathrm{L} \approx 0.11$ is very high and it precisely illustrates the problem already discussed by \citet{Gabriel14}. The introduction of the PM scheme leads to a significant improvement, but it fixes the problem only partially (the second panel of Figure\,\ref{fig_gradients}). The convective core, without inhibited growth, is more massive, with $M_\mathrm{cc}=0.190\,M_\odot$ versus $M_\mathrm{cc}=0.121\,M_\odot$ for the standard mixing, but gradient neutrality on the convective side is not achieved. There is still a very steep decrease of $\nabla_\mathrm{rad}$ from supper- to sub-adiabatic value and a thin semiconvective zone with $\mathrm{d}q \approx 0.002$. Nevertheless, the difference, $\nabla_\mathrm{rad} - \nabla_\mathrm{L} \approx 0.04$ at $q = 0.397$ is lower than in the previous case.

The problem of the gradient inequality is resolved by the CPM scheme. It is visible in the third panel from the top of Figure\,\ref{fig_gradients} that $\nabla_\mathrm{rad}$ gently decreases and the transition from super- to sub-adiabatic is very smooth. The radiative gradient drops below $\nabla_\mathrm{L}$ at $q = 0.308$ and below $\nabla_\mathrm{ad}$ at $q = 0.310$. Once again there is a thin semiconvective zone in the sense of \citet{Kato66}, which is the result of the Ledoux criterion. The most interesting feature of the models with the CPM scheme is the presence of a large semiconvective zone in the sense of \citet{Schwarzschild58} in the region with gradient neutrality \citep{Salaris17}. In the third panel from the top of Figure\,\ref{fig_gradients}, we can see that the Ledoux gradient increases above the core, which is directly related to the gradient of chemical abundance and hence illustrates that this region is partially mixed due to the CPM algorithm. The mass of the convective core is $M_\mathrm{cc}=0.146\,M_\odot$, but the total mass of the mixed region, which combines convective and semiconvective zone, is $M_\mathrm{sc}=0.204\,M_\odot$. The top boundary of the semiconvective region is marked by the steep increase of $\nabla_\mathrm{L}$ at the transition between the mixed area and the uniform helium-rich envelope. In the bottom panel of Figure\,\ref{fig_gradients}, the semiconvective zone is magnified and the behaviour of $\nabla_\mathrm{rad}$ is visible in detail. The value of the radiative gradient is at an approximately constant level above the convective core until it drops in the helium envelope. Throughout most of the semiconvective region $\nabla_\mathrm{rad} <  \nabla_\mathrm{ad}$ and hence the zone is stable versus convection and semiconvection in the sense of \citet{Kato66} and the energy transfer is radiative. The radiative gradient is not perfectly smooth and there are thin layers where its value might be close or slightly higher than the value of $\nabla_\mathrm{ad}$. Such layers are formally semiconvective in the sense of \citet{Kato66}, but they seem not to be important for our models and we do not plot them in the bottom panel of Figure\,\ref{fig_kipp} (Section\,\ref{sec_core}). 

\begin{figure}
  \includegraphics[clip,width=\columnwidth]{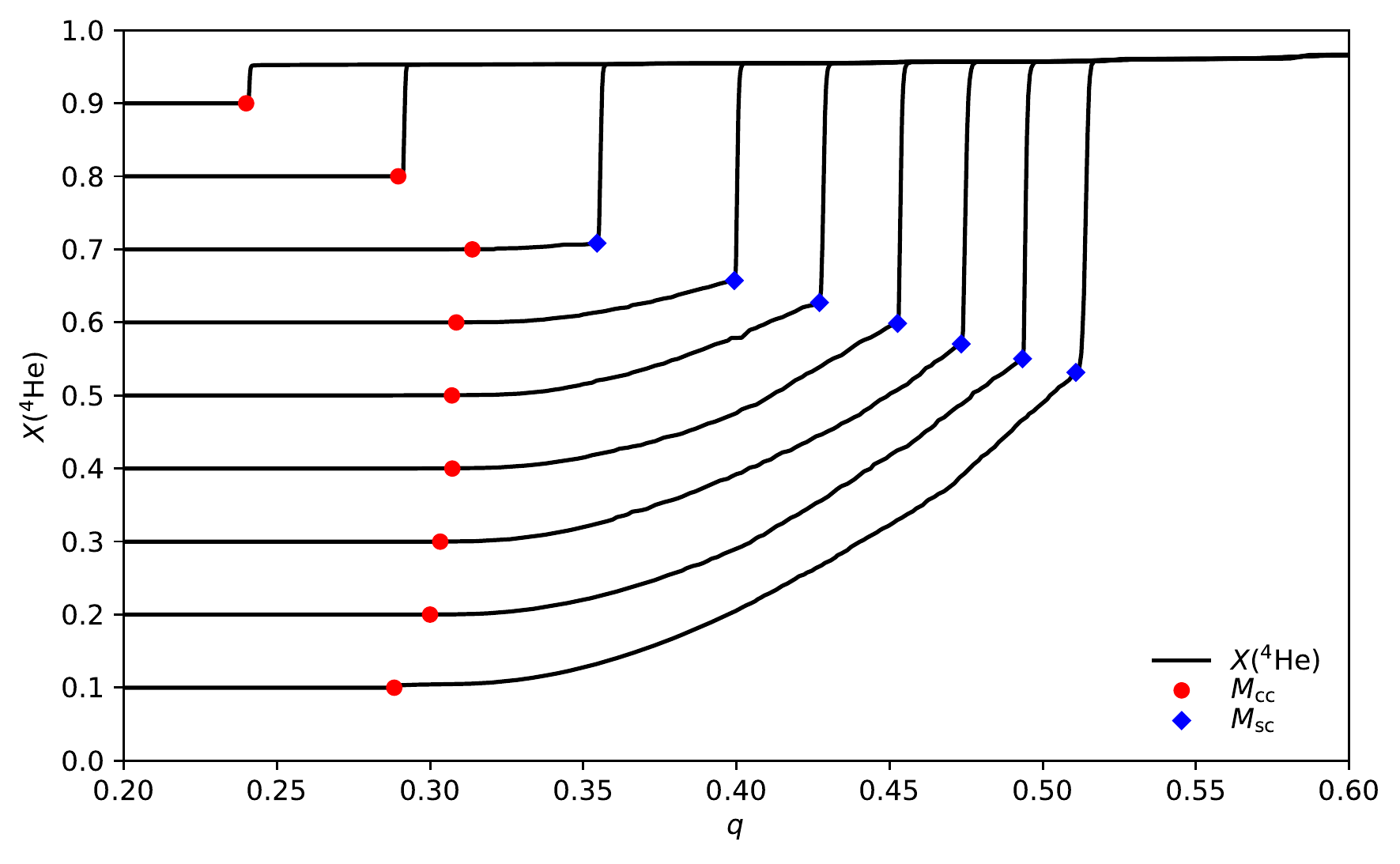}
  \caption{Helium mass fraction, $X(^{4}\mathrm{He})$, in function of relative mass, $q$, for models with the CPM scheme and the central helium abundances in range of $Y_\mathrm{c} = 0.9 - 0.1$. Red dots depict the boundary of convective core and blue diamonds mark the top of the semiconvective region.}
  \label{fig_he4}
\end{figure}

The partial mixing above the convective core in the models with the CPM scheme is illustrated in Figure\,\ref{fig_he4}, where the profile of helium abundance, $X(^{4}\mathrm{He})$, is plotted in function of $q$ for models with the central helium abundances in a range of $Y_\mathrm{c} = 0.9 - 0.1$. The boundary of the convective core is shown with red dots and the blue diamonds mark the top of the semiconvective region. The model with $Y_\mathrm{c} = 0.5$ is the same model as considered in the third panel from the top of Figure\,\ref{fig_gradients}. There is a rather sharp transition between the convective core and the envelope in models with $Y_\mathrm{c} = 0.9$ and $0.8$, in which the semiconvective region \citep[in the sense of][]{Schwarzschild58} has not yet been developed. In the considered model this happens at $Y_\mathrm{c} \approx 0.75$ and we can see the semiconvection in the model with $Y_\mathrm{c} = 0.7$. The size of this region increases with the evolution and in the following models the helium abundance smoothly increases across the partially-mixed zone. In the third panel from the top of Figure\,\ref{fig_gradients}, this smooth change of abundances is visible in the increase of the Ledoux gradient above the convective core. The last zone in the semiconvective region is the last zone mixed by the CPM algorithm. The layers above that are not mixed, except for a very small effect caused by chemical diffusion. Therefore, there is a steep change in the abundances, also visible in Figure\,\ref{fig_gradients} as a sharp peak of $\nabla_\mathrm{L}$ at the top of the mixed region.

The local minimum of $\nabla_\mathrm{rad}$, visible in the PM model in the second panel from the top of Figure\,\ref{fig_gradients}, is known to cause problems with core-helium-burning models if it drops below $\nabla_\mathrm{ad}$ (Section\,\ref{sec_introduction}). In such a case the core splits into a smaller convective core and a convective shell with a radiative region in-between. This is not proper and desirable behaviour \citep[e.g.,][]{Salaris17}. In order to prevent such an event we use the approach similar to the one used by \citet{Constantino15}. If, during the PM iterations, the value of $\nabla_\mathrm{rad}/\nabla_\mathrm{ad} - 1$ drops below the chosen threshold anywhere in the mixed region, then the code backs off the mixing by one cell and updates the convective diffusion coefficient to prevent splitting of the convective zone \citep[][cf. Appendix\,\ref{sec_models}]{Paxton18}. The method works well in most cases, but we have encountered evolved PM models ($Y_\mathrm{c} < 0.3$) in which the core splits despite this safeguard parameter. This artificial modification of models can be avoided by using the CPM scheme, which assures gradient neutrality and hence eliminates the problem entirely. A serious drawback of the CPM scheme is the complete lack of mechanisms preventing breathing pulses. Nevertheless, we consider the achieved gradient neutrality and fewer artificial changes to the models as solid arguments to prefer the CPM scheme over the PM scheme within its range of usability, i. e. for $Y_\mathrm{c} \gtrsim 0.1$.

The local minimum of $\nabla_\mathrm{rad}$ is yet another reason why the standard-mixing model cannot be properly resolved in time. Increasing the temporal resolution in this model allows the convective core to grow faster due to element-diffusion mixing, but it quickly leads to a deep local minimum of $\nabla_\mathrm{rad}$ and the splitting of the core. Further evolution of the convective core is then similar to the evolution of model with standard mixing and convective overshooting (Figure\,\ref{fig_kipp_overshooting}). In that case, the overshooting provides additional mixing above the core, which also leads to a faster growth of the convective core, but also does not have any safeguard mechanism preventing splitting of the core due to the minimum of $\nabla_\mathrm{rad}$. This situation is unrealistic and hence the sdB models with standard mixing or with overshooting from the helium-burning-core should be avoided.

\section{Comparison with asteroseismology} \label{sec_asteroseismology}

\begin{figure}
  \includegraphics[clip,width=\columnwidth]{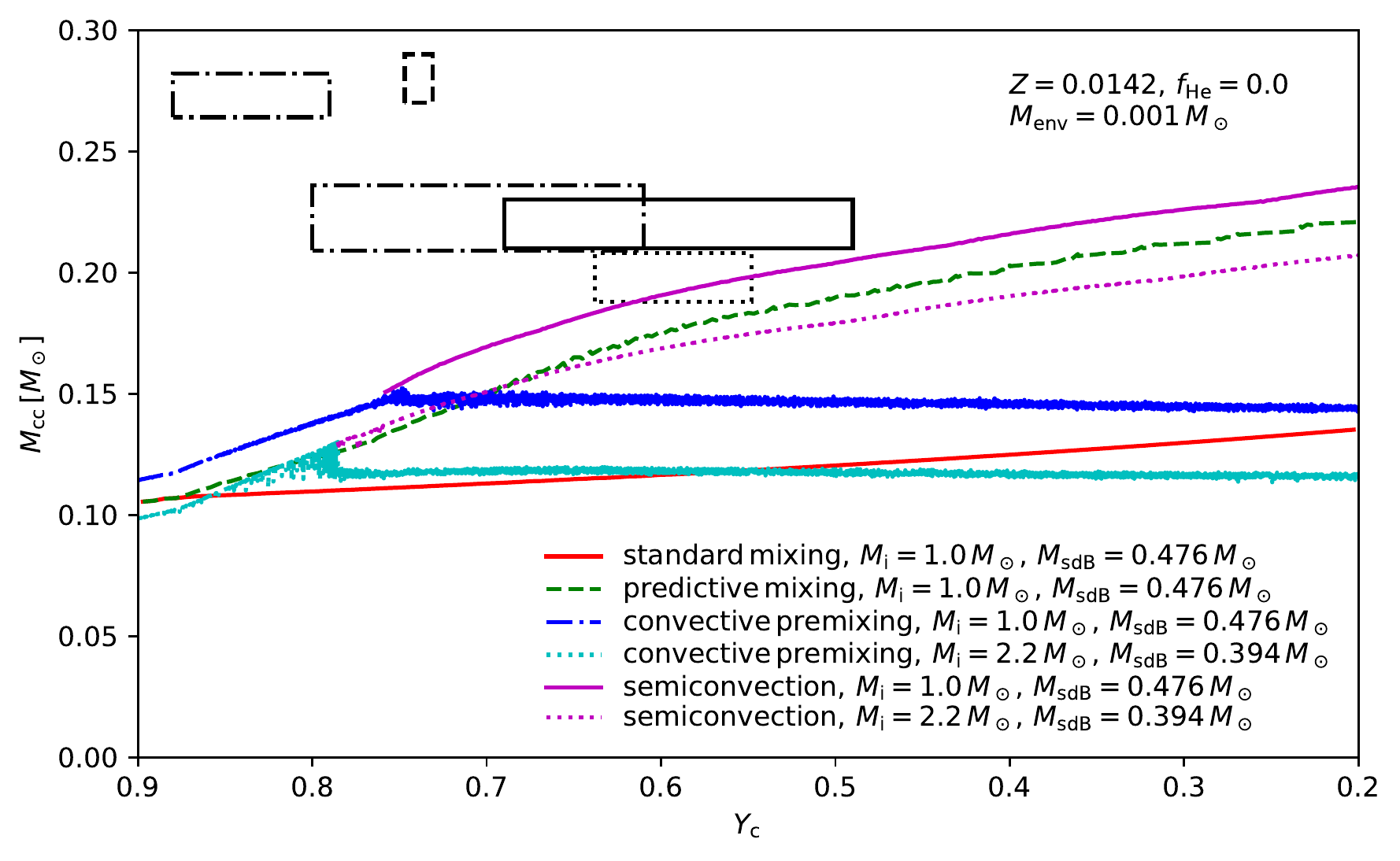}
  \caption{Masses of convective cores, $M_\mathrm{{cc}}$, in function of central helium abundance, $Y_\mathrm{c}$, for the models with metallicity $Z = 0.0142$ and no overshooting from the helium core. Three models from Figure\,\ref{fig_kipp} are shown: the model with standard mixing: solid-red line, the PM scheme: the dashed-green line, the CPM scheme: the dashed-dotted-blue line. The cyan-dotted line shows the model with the CPM scheme and higher mass of progenitor, $M_\mathrm{i} = 2.2 M_\odot$. The magenta lines show the top boundary of the semiconvective zones for the models with the CPM scheme. Asteroseismic results for g-mode pulsators are shown with black colour and with solid-line rectangle: \citet{vanGrootel10a}, dashed-line rectangle: \citet{vanGrootel10b}, dashed-dotted-line rectangles: \citet{Charpinet11}, and dotted-line rectangle: \citet{Charpinet19}.}
  \label{fig_astero_comparison}
\end{figure}

In Section\,\ref{sec_introduction}, we introduced the problem of discrepancy between convective core masses derived from asteroseismology of g-mode pulsators and the results of evolutionary calculations. In Figure\,\ref{fig_astero_comparison}, we compare masses of convective cores, $M_\mathrm{cc}$, with the asteroseismic results shown in black colour and with solid-line rectangle: \citet{vanGrootel10a}, dashed-line rectangle: \citet{vanGrootel10b}, dashed-dotted-line rectangles: \citet{Charpinet11}, and dotted-line rectangle: \citet{Charpinet19}. Age is not a basic parameter of the static models and is also not provided by \citet{Charpinet19}, therefore, the masses are plotted versus central helium abundance, $Y_\mathrm{c}$. We consider the same models as presented in Figure\,\ref{fig_kipp} with a mass of progenitor $M_\mathrm{i} = 1.0\,M_\odot$ and a mass of helium core $M_\mathrm{He\,core} = 0.475\,M_\odot$ and an additional CPM model with a mass of progenitor, $M_\mathrm{i} = 2.2\,M_\odot$, and $M_\mathrm{He\,core} = 0.393\,M_\odot$.

The model with standard mixing (red-solid line) illustrates the essence of the core-mass-discrepancy problem. The mass of the convective core in the suitable range of $Y_\mathrm{c} \approx 0.88 - 0.49$ is $M_\mathrm{cc} \approx 0.10 - 0.12\,M_\odot$, which is lower by about $0.10 - 0.18\,M_\odot$ than the seismic-derived masses. The difference is significant and clearly shows that the standard models of sdBs are inadequate.

In the case of the PM scheme (green-dashed line), the convective core grows from about $0.105$ to $0.190\,M_\odot$ in the range of $Y_\mathrm{c} \approx 0.88 - 0.49$. For the model with the CPM scheme and $M_\mathrm{sdB} = 0.476\,M_\odot$, the mixed region, in the same range of $Y_\mathrm{c}$, grows from $M_\mathrm{cc} = 0.117\,M_\odot$ (convective core; blue-dashed-dotted line) to $M_\mathrm{sc} = 0.205\,M_\odot$ (semiconvective region; magenta-solid line). This is a visible increase of $M_\mathrm{cc}$ versus the standard model, but none of the models can reach the same extend of the mixed region as obtained by \citet{vanGrootel10b} for KPD 1943+4058 ($M_\mathrm{cc} = 0.28 \pm 0.01\,M_\odot$), \citet{Charpinet11} for KIC 02697388 ($M_\mathrm{cc} = 0.274^{+0.008}_{-0.010}\,M_\odot$ or $M_\mathrm{cc} = 0.225^{+0.011}_{-0.016}\,M_\odot$), and \citet{vanGrootel10a} for KPD 0629-0016 ($M_\mathrm{cc} = 0.22 \pm 0.01\,M_\odot$).

The three g-mode pulsators of \citet{vanGrootel10a,vanGrootel10b} and \citet{Charpinet11} have total masses close to the canonical value, $M_\mathrm{sdB} = 0.452 - 0.496\,M_\odot$, while the star recently studied by \citet[EC 21494-7018]{Charpinet19} has a mass of $M_\mathrm{sdB} = 0.391 \pm 0.009\,M_\odot$ and the core mass $M_\mathrm{cc} = 0.198 \pm 0.010\,M_\odot$. The previously analysed models with total mass $M_\mathrm{sdB} = 0.476\,M_\odot$ are not suitable for this object and hence we calculated the more adequate CPM model with mass $M_\mathrm{sdB} = 0.394\,M_\odot$. When this model is considered the mass of the partially-mixed region (magenta-dotted line) is once again too low. The mass of the semiconvective region is lower by about $0.02\,M_\odot$ as compared the more massive CPM model.

We can see that none of the mixing schemes available in \texttt{MESA} can solve the core-mass-discrepancy problem, even if the definition of core is extended from the convective region to the partially-mixed region. It might seem that one-dimensional evolution codes are unable to solve the problem. The PM scheme, similarly to the maximal overshoot scheme of \citet{Constantino15} that inspired the algorithm, produces the largest possible convective core during the core-helium-burning evolution. The expanse of the partially-mixed zone in the models with the CPM scheme is systematically even larger, but still too small for the seismic-derived values. There is no significant improvement versus the old models calculated by \citet{Sweigart87} with a large semiconvective zone \citep[cf. Figure\,3 in][]{Schindler17}. However, the PM and CPM schemes yield important improvement versus the previous \texttt{MESA} models \citep{Schindler15}, as they allow for significant growth of the central mixed region.

\section{Period spacing} \label{sec_periods}

\begin{figure}
  \includegraphics[clip,width=\columnwidth]{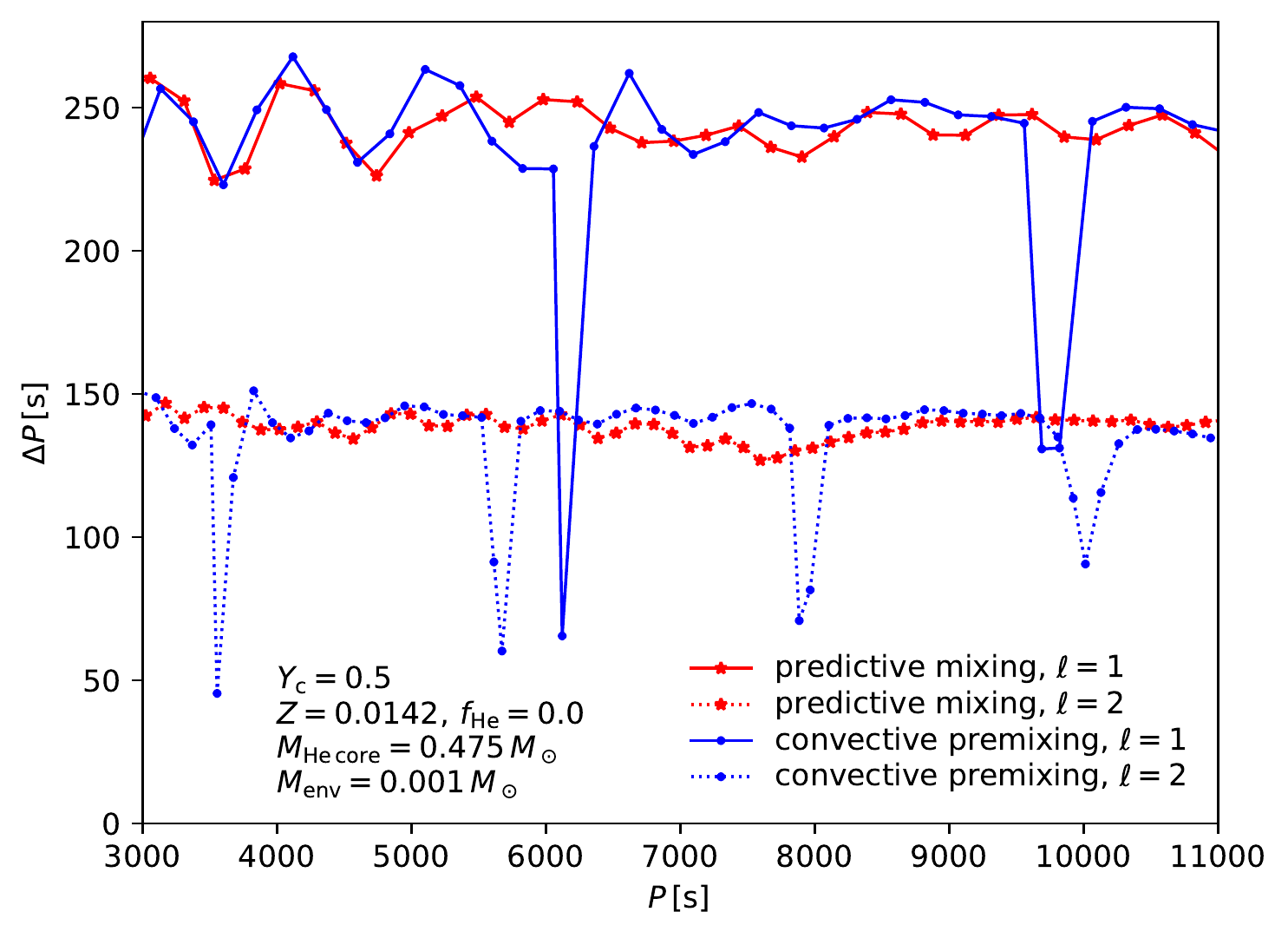}
  \includegraphics[clip,width=\columnwidth]{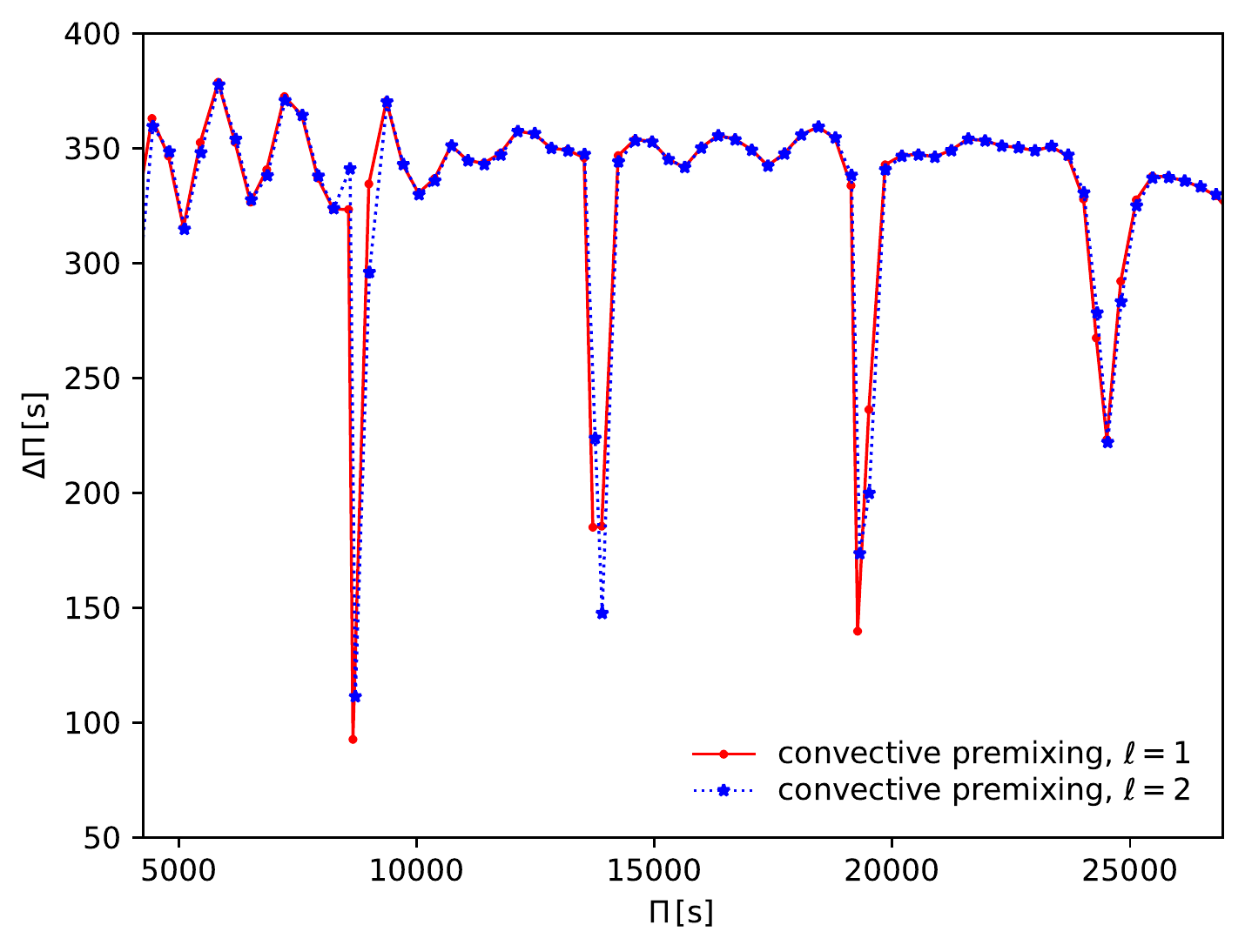}
  \caption{Top panel: period spacing, $\Delta P$, versus period, $P$, for the models with the PM scheme (red lines and stars) and the CPM scheme (blue lines and dots), and central helium abundance, $Y_\mathrm{c} = 0.5$ (same models as shown in Figure\,\ref{fig_gradients}). The modes of degrees $\ell=1,2$ are shown with the solid and dotted lines, respectively. Bottom panel: reduced period spacing, $\Delta \Pi$, versus reduced period, $\Pi$, for the CPM model. Red solid line with dots and blue dotted line with stars depict the modes of degrees $\ell=1,2$, respectively.}
  \label{fig_period_spacing}
\end{figure}

An additional comparison between models with the PM scheme and the CPM scheme can be performed by utilizing the period spacing of gravity modes. In this paper we are not focused on pulsations and their properties, but period spacings are very easy to calculate and they provide a valuable insight into the structure of the considered models.

The simple adiabatic pulsation models for selected \texttt{MESA} models are calculated using the publicly available oscillation code GYRE, version 5.2 \citep{Townsend13,Townsend18}. The asymptotic relations for high-order g-modes predict that for a given modal degree, $\ell$, the periods of modes with consecutive radial orders, $n$, should be equidistant \citep{Tassoul80}. This is visible in the top panel of Figure\,\ref{fig_period_spacing}, in which period spacing, defined as $\Delta P = P_{n,\ell} - P_{n-1,\ell}$, is plotted as a function of period, $P$, for the models with the PM scheme (red lines and stars) and the CPM scheme (blue lines and dots). The models have central helium abundance, $Y_\mathrm{c} = 0.5$ (the same as in models shown in Figure\,\ref{fig_gradients}). The modes of degrees $\ell=1,2$ are shown with the solid and dotted lines, respectively.

The period range considered in the top panel of Figure\,\ref{fig_period_spacing}, $P = 3000$ - $11000$ seconds, is chosen to roughly correspond to the periods detected in the Kepler data \citep{Reed18}. The radial orders in the period range considered are $n = [12, 44]$ for $\ell = 1$, $n = [21, 78]$ for $\ell = 2$ for the PM model, and $n = [13, 46]$ for $\ell = 1$ and $n = [22, 81]$ for $\ell = 2$ for the CPM model.

The period spacing for the modes of degree $\ell = 1$ is about $250$ seconds and for the modes of degree $\ell = 2$ about $150$ seconds, for both mixing prescriptions. This is within the expected range of values for the pulsating sdB stars \citep{Reed18}. While the average $\Delta P$ is similar in both algorithms, there is one immediately visible difference: the model with the PM scheme does not have any trapped modes within the given range, whereas the trapping is present in the model with the CPM scheme. Trapping can occur if there are sharp features in the chemical composition and the Brunta-V\"{a}is\"{a}la frequency \citep[e.g.][]{Unno89,Paxton13} across the stellar structure that may lead to emerging of additional acoustic cavities \citep{Dziembowski93,Miglio08,Cunha15}. Examples and discussion on mode trapping in pulsating sdBs can be found in, e.g., \citet{Charpinet00,Charpinet02a,Charpinet02b,Ghasemi17}.

Reduced period diagram presents reduced period spacing, $\Delta \Pi = \sqrt{\ell(\ell + 1)} \Delta P$, in function of reduced period, $\Pi = \sqrt{\ell(\ell + 1)} P$. It is one of the best tools to look for trapped modes, because the multiplication causes the sequences of modes with all degrees to overlap \citep[e.g.,][]{Ostensen14,Baran17,Sahoo20}. Such a diagram is shown in the bottom panel of Figure\,\ref{fig_period_spacing} and it clearly shows that the modes of degrees $\ell = 1$ and $2$ overlap and that the trapping is indeed present in the CPM model.

Comparison of the internal structure of the models can easily explain the presence or the lack of trapped modes in the considered oscillation models. As previously discussed in Sections\,\ref{sec_core} and \ref{sec_gradients}, the internal structure of the PM model is rather simple with the fully-mixed convective core sharply transitioning into the helium-rich envelope. With such configuration, the pulsation modes have to be reflected at the border of the convective core. In the case of the convective premixing model, the partially mixed zone exists between the convective core and the helium envelope. Some modes can be trapped in the additional acoustic cavity in the semiconvective region, which explains the differences in Figure\,\ref{fig_period_spacing} \citep[e.g.][]{Unno89,Charpinet00,Daszynska13}. We consider the presence of the trapped modes as another argument for the CPM scheme.

\section{Summary and conclusions} \label{sec_summary}

In this paper we have explored the properties of \texttt{MESA} models of sdB stars, calculated with the new algorithms for treatment of convective boundaries: the predictive mixing and, for the first time, the convective premixing scheme. Our goals were to obtain models with a smooth, reliable internal structure and correctly determined boundaries of the convective core, free from the problems related to the core-helium-burning, and with cores large enough to be comparable with the seismic-derived results.

The \texttt{MESA} models of sdB stars calculated in past were limited only to the Ledoux or Schwarzschild criteria for convection and produced masses of the convective cores significantly lower than predicted by the static asteroseismic models \citep{vanGrootel10a,vanGrootel10b,Charpinet11,Charpinet19}. In Section\,\ref{sec_core}, we showed for the representative models, that the PM scheme allows for a significant growth of the convective core when compared with models using only the Ledoux criterion. Nevertheless, it does not fully solve the issue of determining the convective boundary discussed by \citet{Gabriel14}. This problem is solved by the CPM scheme, which achieves gradient neutrality, $\nabla_\mathrm{rad} \simeq \nabla_\mathrm{ad}$, at the boundary of the convective core. In models with the CPM scheme, the convective core is smaller than in the PM models, but the large partially-mixed semiconvective zone emerges in the region with gradient neutrality. In this case we no longer treat the core as a fully-mixed, convective region, but we extend the definition to the region that consists of the fully-mixed convective zone and the partially-mixed semiconvective zone \citep[in the sense of][]{Schwarzschild58}. The semiconvection occurs as an effect of the CPM algorithm and does not have to be introduced by a specific prescription \citep[e.g.,][]{Constantino15}.

The internal structure of the sdB models calculated with both algorithms and no overshooting from the convective core is smooth. In the case of the PM scheme, there are methods that suppress breathing pulses and prevent core splitting in most of the models. The problem of splitting of the convective core due to the local minimum of $\nabla_\mathrm{rad}$ does not occur in the models with the CPM scheme, due to the always achieved gradient neutrality. However, models with this algorithm are usable only until the central helium abundance drops to about $Y_\mathrm{c} \approx 0.1$. At this evolutionary stage, the breathing pulses occur and the sudden increase of the convective core destroys the gentle abundance gradient in the semiconvective zone. We treat this behaviour as a numerical artefact and hence do not consider the models after the onset of the first breathing pulse as realistic. Currently there is no safeguard against breathing pulses in the CPM scheme and they occur in every calculated model.

In Section\,\ref{sec_overshooting}, we found that convective overshooting does not work well during the core-helium-burning phase and with both discussed algorithms. Because of the numerous issues and the lack of additional growth of the mixed region we do not recommend including overshooting when calculating sdB \texttt{MESA} models. 

Unfortunately, none of the models that we presented is able to reproduce the core masses derived from asteroseismology. There seems to be no method of increasing the mixed region further in the \texttt{MESA} models, which might be a limitation of the one-dimensional approach to modelling of convection and other mixing mechanisms. There is also a possibility that future improvements to the static asteroseismic models would lower the obtainable core masses.

The most significant advantages of the CPM scheme over the PM scheme are solving the problem of determining the convective boundaries by achieving gradient neutrality, no need of introducing artificial parameters to prevent the splitting of convective core, obtaining a larger mixed region and the presence of trapped modes in the oscillation spectra after the emergence of the semiconvective zone. The biggest drawback of this algorithm is usability limitation to the point when $Y_\mathrm{c} \approx 0.1$. That also prevents all studies of post-sdB objects. Nevertheless, due to the important advantages, we intend to use the CPM models in future work and comparison to the asteroseismic targets. The lifetime of the CPM model to the first breathing pulse is about $80\%$ of the predictive mixing model lifetime. Therefore, it is possible to work within the usable range of $Y_\mathrm{c}$ for most of the observable sdB stars.

This paper also creates a foundation for our future work. Models with the setup and properties presented here allow for calculations of adiabatic pulsation models and hence comparison with numerous pulsating hot subdwarfs observed during Kepler and TESS missions. In the future, we plan to extend this work by including radiative levitation and to focus more on the pulsation-driving regions so that the models would be suitable for non-adiabatic pulsation calculations.

\section*{Acknowledgements}

We gratefully thank Evan Bauer, Anne Thoul and Josiah Schwab for helpful comments and advices, and Katarzyna Sta\'{n}ko-Ostrowska for numerous language corrections.

This work was financially supported by the Polish National Science Centre grants UMO-2017/26/E/ST9/00703 and UMO-2017/25/B/ST9/02218.

Calculations have been carried out using resources provided by Wroclaw Centre for Networking and Supercomputing (http://wcss.pl), grant No. 265.

\textit{Software}: \texttt{MESA SDK} version 20190503 \citep{Townsend19}, \texttt{Python} (\href{https://www.python.org)}{https://www.python.org}), \texttt{PyMesaReader} \citep{Wolf17a}, \texttt{iPython} \citep{Perez07}, \texttt{jupyter} \citep{Kluyver16}, \texttt{matplotlib} \citep{Hunter07}, \texttt{NumPy} \citep{vanderWalt11}, \texttt{Pandas} \citep{McKinney10}, \texttt{Ruby} (\href{https://www.ruby-lang.org}{https://www.ruby-lang.org}), \texttt{MesaScript} \citep{Wolf17b}, \texttt{Kippenhahn plotter for MESA} \citep{Marchant19}.

\section*{Data availability}

The data underlying this article will be shared on reasonable request to the corresponding author.

\appendix

\section{Convective boundaries} \label{sec_convective_boundaries}

The standard way to locate the convective boundaries is using the Schwarzschild criterion, $y = \nabla_\mathrm{rad} - \nabla_\mathrm{ad}$, or Ledoux criterion $y = \nabla_\mathrm{rad} - \nabla_\mathrm{L}$, where $\nabla_\mathrm{rad}$ is the radiative gradient, $\nabla_\mathrm{ad}$ the adiabatic gradient and $\nabla_\mathrm{L}$ is the Ledoux gradient \citep{Paxton13}. The convective boundary is found when the discriminant $y$ changes its sign. \citet{Gabriel14} argued that this approach leads to problems if the criterion $\nabla_\mathrm{rad} = \nabla_\mathrm{ad}$ does not hold on the convective side. This is because the convective boundary should be defined as a point when the convective velocity vanishes and the MLT is well defined only on the convective side. This scheme works if the chemical composition is continuous across the convective boundary, but this is rarely the case in stellar models. The composition discontinuity leads to discontinuities in density and opacity and hence to discontinuities of $\nabla_\mathrm{rad}$, $\nabla_\mathrm{ad}$ and  $\nabla_\mathrm{L}$. In such situation $\nabla_\mathrm{rad} > \nabla_\mathrm{ad}$ on the convective side and the growth of the convective region is prevented. According to \citet{Gabriel14} this behaviour is not physical and not in agreement with MLT. The application to models of sdB stars of the two mechanisms available in \texttt{MESA} that are supposed to fix this problem, the predictive mixing scheme and the convective premixing scheme, is the main subject of this paper.

\subsection{Predictive mixing} \label{sec_predictive}

The \texttt{MESA} predictive mixing scheme was introduced in \citet{Paxton18}. It was influenced by previous attempts to solve the convective boundary problem, especially by the maximal overshoot scheme of \citet{Constantino15} and by the procedure from \citet{Bossini15}. The algorithm starts with finding the cells where $y$ changes its sign. Then, it mixes the first cell on the radiative side (candidate cell) so it would have the same composition as the adjacent cell on the convective side. The parameters such as opacity or density are adjusted accordingly. If $y$ would become positive on both sides of the convective boundary the next cell becomes the candidate cell. The algorithm proceeds until the candidate cell still has negative $y$ after the proposed mixing. Subsequently, the convective velocities and diffusive parameters are recalculated using the MLT and committed to the model. The PM scheme does not directly change the chemical composition. All details of the procedure can be found in \citet{Paxton18}.

The physical justification of the PM scheme is based on the assumption of \citet{Castellani71} that any gentle mixing outside the helium-burning core changes the composition there and hence changes the opacity and gradients leading to a gradual increase of the core. The nature of this mixing is irrelevant. It is shown in this paper that the PM scheme is useful, but it does not solve all the problems with the convective boundaries.

\subsection{The convective premixing scheme} \label{sec_premixing}

The second algorithm available in \texttt{MESA}, originally meant to solve the shortcomings of the PM scheme, is the convective premixing scheme, introduced by \citet{Paxton19}. Again, all details of the implementation can be found in the original paper, here we only explain the basics briefly.

In a similar fashion to the PM scheme, the algorithm starts with finding cells where the discriminant $y$ changes its sign. It is done at the beginning of the time step, before other changes are applied. Then, it checks if the sign of $y$ of the candidate cell on the radiative side would change if the cell is mixed completely with the rest of the convection region. The tentative mixing is performed with pressure, temperature, abundances, densities, opacities and the temperature gradients recalculated through the candidate cell and the adjacent convective region. If the candidate cell becomes convective the mixing is instantly committed to the model. After that, the next adjacent radiative cell becomes the candidate cell and the iterations continue until the radiative cell outside the current convective boundary remains radiative during the tentative mixing.

The actual mixing of matter while the algorithm is working is the most substantial change in the CPM scheme versus the PM scheme, which only modifies the MLT diffusion coefficients. It leads to the emergence above the core of a partially mixed zone  that increases the size with the evolution during the core-helium-burning phase. This region is a semiconvective zone in the sense of \citet{Schwarzschild58}, which is a different phenomenon than the semiconvection in the sense of \citet{Kato66} that requires the Ledoux criterion for convective instability and is usually considered in the studies of stellar structure and evolution. The CPM scheme can be regarded as a \texttt{MESA} implementation of induced semiconvection \citep{Castellani71,Mowlavi94,Constantino15}.

\subsection{Breathing pulses} \label{sec_breathing}

At the end of the core-helium-burning phase the so-called breathing pulses may emerge. This is a long-known and questionable phenomenon that can occur in models of low-mass stars, both sdBs and regular horizontal-branch stars with a massive envelope \citep{Sweigart73,Castellani85}. When the central helium abundance drops to about $Y_\mathrm{c} \approx 0.1$, $\alpha$-captures by $^{12}$C nuclei dominate the carbon production over the $3\alpha$ reaction. The helium burns mainly due to the $^{12}$C $+$ $^{4}$He $\rightarrow$ $^{16}$O $+$ $\gamma$ reaction. This leads to a fast increase of oxygen abundance, which has higher opacity than carbon, $^{12}$C. As a consequence of higher opacity and issues with determination of the correct boundary of the core, it is plausible that helium from outside the core is ingested into the near-depleted core. At this point of the evolution, even a small amount of fresh helium added to the core enhances the rate of energy production and hence the increase of luminosity and radiative gradient, $\nabla_\mathrm{rad}$. This leads to sudden growth of the convective core and the emergence of a breathing pulse. Then the helium is burnt in the core and the star re-adjusts its structure. Next breathing pulses may follow and usually, when they occur, there are a few of them before the helium is entirely depleted in the core and the evolution continues to the next stage.

There are a few evolutionary effects of the breathing pulses: at each pulse a model performs a loop in the H-R diagram, the He-burning lifetime is increased, and the mass of the CO-core at helium exhaustion is increased \citep{Salaris17}. The existence of breathing pulses is debatable and they are most probably numerical artefacts and not the real features of the stellar evolution \citep{Caputo89,Dorman93,Cassisi01,Cassisi03,Constantino16,Constantino17}.

\section{Evolutionary models} \label{sec_models}

Evolutionary models are calculated using publicly available and open source code \texttt{MESA} \citep[Modules for Experiments in Stellar Astrophysics;][]{Paxton11,Paxton13,Paxton15,Paxton18,Paxton19}, version 11701. Here, we discuss the physical and numerical setup used in the models of the subdwarfs.\footnote{Detailed settings and sample scripts are provided online: \href{https://github.com/cespenar/grid_sdb}{\url{https://github.com/cespenar/grid_sdb}}} We use the values of parameters provided below in all models, unless explicitly stated otherwise.

We adopt OPAL opacity tables \citep{Iglesias96} extended by the data of \citet{Ferguson05} for the lower temperatures and of \citet{Buchler76} for Compton-scattering dominated regime. Electron conduction opacities are from \citet{Cassisi07}. During more advanced phases of the evolution, type 2 opacities were used to account for varying amounts of carbon and oxygen beyond that accounted for by $Z$. We use the standard \texttt{MESA} equation of state (EOS), which is a blend of the OPAL \citep{Rogers02}, SCVH \citep{Saumon95}, PTEH \citep{Pols95}, HELM \citep{Timmes00}, and PC \citep{Potekhin10} EOSes. The adopted nuclear reaction rates are taken from the JINA REACLIB database \citep{Cyburt10,Rauscher00} supplemented by additional tabulated weak reaction rates \citep{Fuller85,Oda94,Langanke00}. Screening is included via the prescription of \citet{Chugunov07} and thermal neutrino loss rates are from \citet{Itoh96}. The JINA REACLIB database contains the most important rates for the helium-burning phase in their up-to-date versions: $^{12}$C($\alpha, \gamma$)$^{16}$O \citep{Hammer05}, $^{14}$N(p,$\gamma$)$^{15}$O \citep{Imbriani05} and $3\alpha$ \citep{Fynbo05}. We use a custom reaction network based on the available network \texttt{pp\_cno\_extras\_o18\_ne22}, but with addition of $^{56}$Fe and $^{58}$Ni.

The outer boundary conditions are calculated using the \texttt{grey\_and\_kap}, which iterates a simple grey atmosphere to find consistent pressure, temperature, and opacity at the surface. 

In sdB models element diffusion is enabled according to the formalism described in \citet{Paxton18}. We use eleven classes of species represented by the following isotopes: $^{1}$H, $^{3}$He, $^{4}$He, $^{12}$C, $^{14}$N, $^{16}$O, $^{20}$Ne, $^{22}$Ne, $^{24}$Mg, $^{56}$Fe, and $^{58}$Ni. The radiative levitation was omitted in this paper because it has negligible effects on cores of evolutionary models, which are the focus of this paper. Its effects are critically important for non-adiabatic pulsations \citep[e.g.,][and references within]{Bloemen14} and will be discussed elsewhere. 

Locations of convective zones are determined by the Ledoux criterion for convective instability. In regions stable according to the Ledoux criterion for convection, but unstable according to the Schwarzschild criterion, semiconvective mixing with the scheme of \citet{Langer83} is used. We adopt value of the efficiency parameter $\alpha_\mathrm{sc}=0.1$. We use mixing length formalism of \citet{Cox68} with the mixing-length parameter fixed to $\alpha_\mathrm{MLT}=1.80$. In models with overshooting from the convective regions the exponential formula of \citet{Herwig00} is used:

\begin{equation} \label{eq_overshooting}
	D_\mathrm{ov} = D_\mathrm{conv}\exp\left(-\frac{2z}{fH_P}\right),
\end{equation}

\noindent where $D_\mathrm{conv}$ is the diffusion coefficient derived from the MLT at a user-defined location in the convective zone ($f_0 H_P$ off the boundary of a convective zone; $f_0=0.002$ in all models considered), $H_P$ is the pressure scale height at that location, $z$ is the distance in the radiative layer away from that location and $f$ is an adjustable parameter. Parameter $f$ is different than usually used $\alpha_\mathrm{ov}$ from the step overshooting description and it approximately follows the relation $f \approx 0.1\alpha_\mathrm{ov}$ \citep[e.g.][]{Moravveji15,Ostrowski17,Valle17,Claret17}. 

In order to resolve the problem with proper determination of convective boundaries (Appendix\,\ref{sec_convective_boundaries}), in some models we use more sophisticated algorithms in addition to Ledoux criterion: the PM scheme \citep{Paxton18,Paxton19} or the CPM scheme \citep{Paxton19}, briefly described in Appendices\,\ref{sec_predictive} and \ref{sec_premixing}. If the PM scheme is used, it is enabled only for the convective core. In order to prevent core splitting and breathing pulses, we use \texttt{predictive\_superad\_thresh = 0.05} and \texttt{predictive\_avoid\_reversal = 'he4'}. In models with the CPM scheme enabled core splitting does not occur due to the nature of the algorithm, but there is currently no method of preventing the breathing pulses (Sections\,\ref{sec_core} and \ref{sec_gradients}).

Enabling the CPM scheme requiers very high spatial, \texttt{mesh\_delta\_coeff = 0.2}, and temporal resolution, \texttt{max\_years\_for\_timestep = 20000 yr}.

The rotation is omitted in the calculated sdB models due to their very low measured rotational velocities \citep{Reed18,Charpinet18}. We also neglected the effects of mass-loss and stellar winds. The chemical composition of the sdB models is based on the composition of progenitors (Appendix\,\ref{sec_progenitors}).

\section{Progenitors} \label{sec_progenitors}

In order to obtain a sdB star model, we evolve a star from pre-main-sequence (PMS) phase to the tip of the red giant branch (RGB). Then we remove the outer envelope of a red giant, leaving only the helium core and a small hydrogen envelope on top of it. The envelope is removed using \texttt{relax\_mass}. The procedure simulates sdB stars that evolve through the common envelope scenario or Roche lobe overflow channel in binary systems \citep{Han02}. The applied mass removal is a quasi-static process and as such it does not properly reflect the complicated three-dimensional hydrodynamical behaviour of real systems. We also assume that there are no prior episodes of mass transfer between the components of binary systems before the tip of the red giant branch so that we can evolve a single star. The simplifications are very significant, but we focus on the evolution of sdB stars and not on the exact details of their past evolution prior to the extreme horizontal branch. This approach was earlier adopted by many other authors studying sdB stars, e.g., \citet{Han02}, \citet{Hu09}, \citet{Ostensen12}, \citet{Bloemen14}, \citet{Schindler15}, \citet{Ghasemi17} or \citet{Xiong17}.

\subsection{Setup of models} \label{sec_progenitors_setup}

The models of progenitors are calculated with physics very similar to that described in Appendix\,\ref{sec_models}, but with a few modifications. The outer boundary condition is calculated using pre-calculated tables based on model atmospheres. We use \texttt{photosphere\_tables}, which consists of tables constructed with the PHOENIX code \citep[][$-0.5<\log g<5.5$, $2000<T_\mathrm{eff}<10000$\,K]{Hauschildt99a,Hauschildt99b} and supplemented with the tables of \citet{Castelli03} for higher temperatures ($0.0<\log g<5.0$, $3500<T_\mathrm{eff}<50000$\,K). Element diffusion is enabled during main-sequence (MS) evolution, but the adopted treatment is less advanced than during the subdwarf phase. We used the formalism of \citet{Thoul94} with the default five classes of species ($^{1}$H, $^{3}$He, $^{4}$He, $^{16}$O, $^{56}$Fe). It is a significant simplification versus our treatment of diffusion in sdB models, but the effects of MS diffusion are not critical for the presented results. 

The models thoroughly discussed in this paper have the progenitor with solar abundance, $Z = 0.0142$, $Y = 0.2703$ \citep{Asplund09}, but the conclusions are representative and apply to a wide range of initial metallicities, $Z = 0.001 - 0.035$. The sdB stars are found in various stellar populations, in the Galactic field \citep{Edelmann03,Hirsch08}, in open clusters \citep{Kaluzny93} and in globular clusters \citep{Moehler01,Moni08} and the considered range of $Z$ cover most of the possible cases. The initial helium abundances, $Y$, are calculated using the helium enrichment law, $\Delta Y/\Delta Z = 1.5$ \citep[e.g.,][]{Choi16}, and the initial hydrogen abundances follow the standard formula $X=1-Y-Z$. The mixture of the heavier elements within $Z$ is adopted from \citet{Asplund09}.

\subsection{Helium-flash and subflashes} \label{sec_progenitors_flash}

\begin{figure*}
  \includegraphics[clip,width=\textwidth]{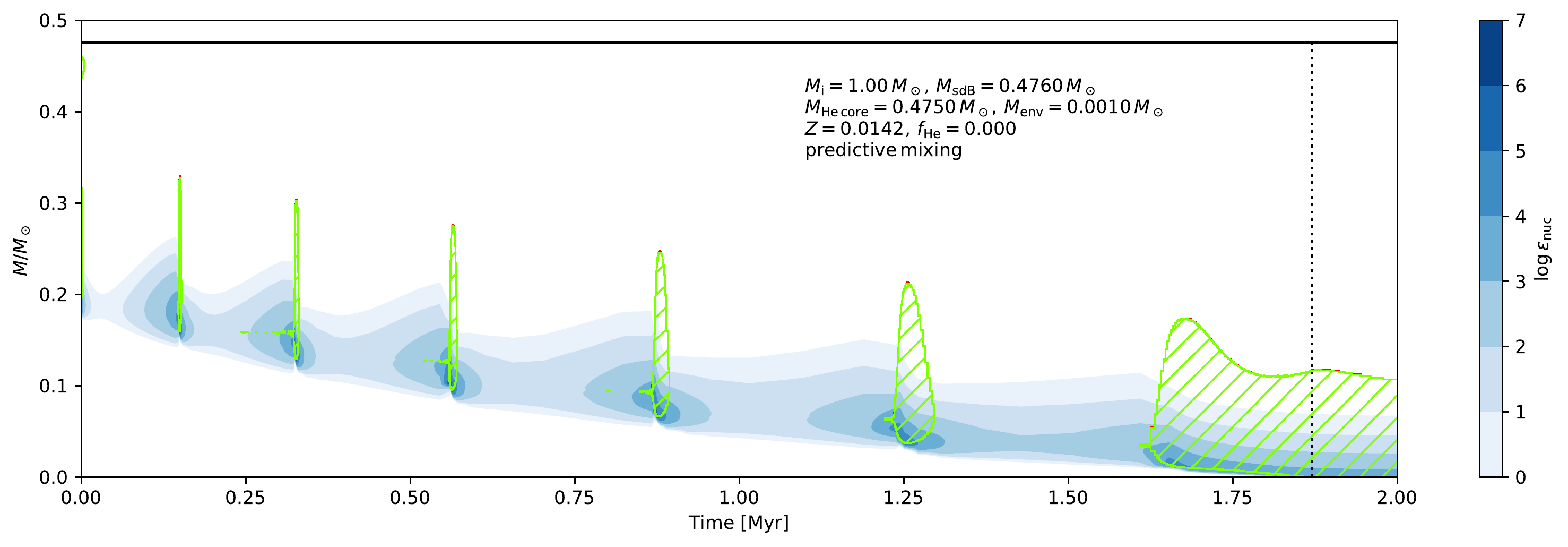}
  \caption{Kippenhahn diagram showing the phase of helium subflashes for a representative sdB model with metallicity $Z=0.0142$, mass of helium core $M_\mathrm{He\,core} = 0.475M_\odot$ and envelope mass $M_\mathrm{env} = 0.001M_\odot$. The PM scheme is used. The green hatched lines depict convective zones, the red hatched lines show semiconvective zones in the sense of \citet{Kato66}. The vertical dotted line marks the moment when the convective zone reaches the centre. The rate of nuclear energy generation, $\log\,\epsilon_\mathrm{nuc}$, is shown in shades of blue. The structure of the star is shown in a function of time elapsed since the removal of the outer envelope.}
  \label{fig_kipp_subflashes}
\end{figure*}

\begin{figure}
  \includegraphics[clip,width=\columnwidth]{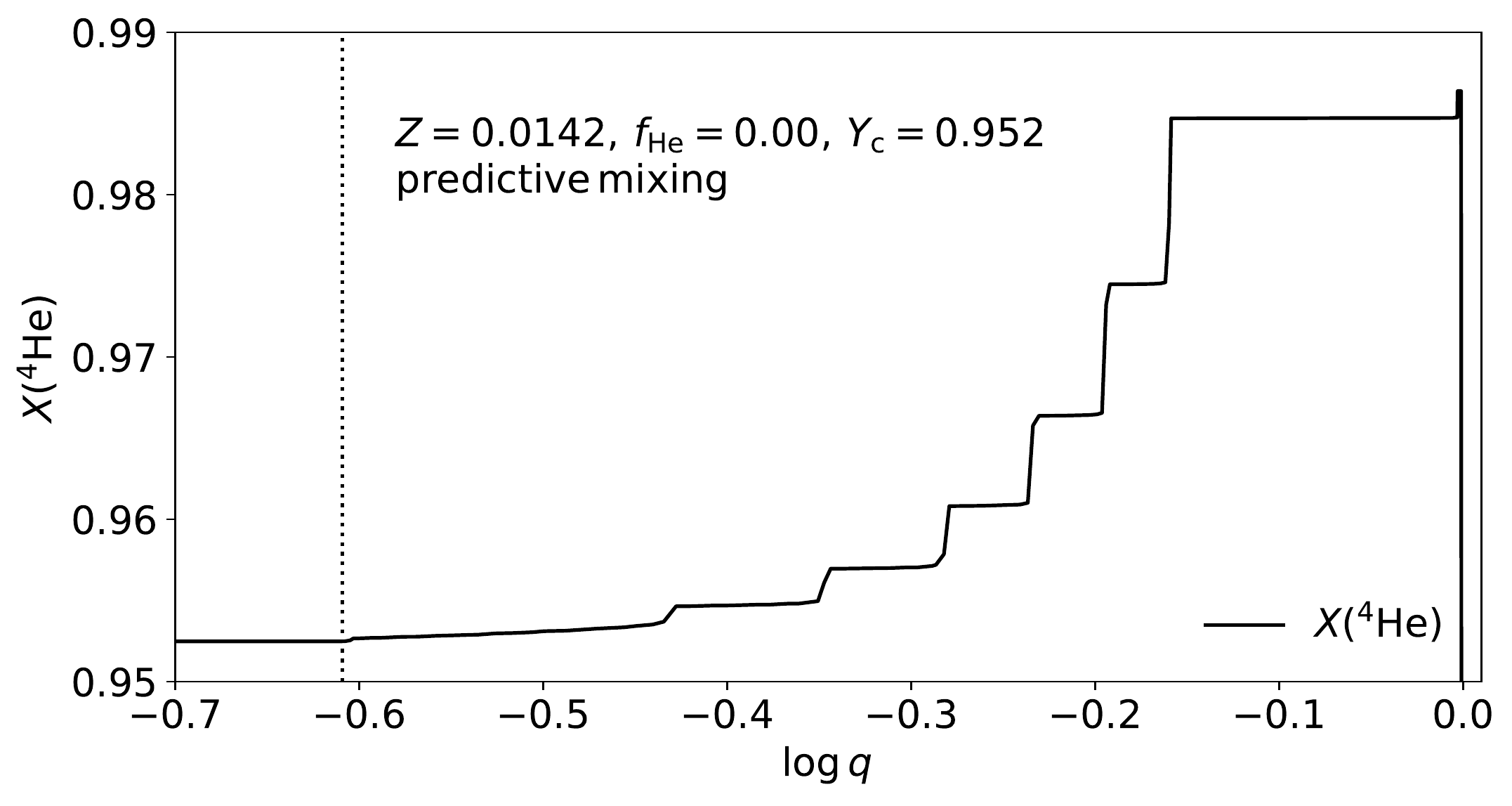}
  \caption{Helium mass fraction, $X(^{4}\mathrm{He})$, in function of logarithmic relative mass, $\log\,q = \log(m/M)$, for the Zero Age Extreme Horizontal Branch model marked by the vertical line in Figure\,\ref{fig_kipp_subflashes}. Only the envelope is shown. The vertical dotted line shows the boundary of the convective core.}
  \label{fig_he4_envelope}
\end{figure}

Stars with masses $M \lesssim 2.0\,M_\odot$ that develop a degenerated core during the RGB evolution ignite helium in an off-center helium flash followed by a series of subflashes that lift the degeneration \citep{Thomas67,Mocak08,Bildsten12,Miller20}. In the models of sdB stars the helium flash occurs after the removal of the envelope. This is illustrated in the Kippenhahn diagram in Figure\,\ref{fig_kipp_subflashes}, which shows the evolution of a representative sdB model with metallicity $Z=0.0142$, mass of helium core $M_\mathrm{He\,core} = 0.475M_\odot$, envelope mass $M_\mathrm{env} = 0.001M_\odot$ and the PM scheme, between the removal of the envelope and the stable core-helium-burning phase. Consecutive subflashes occur in the shells that move inward in mass coordinate and the convective zone, shown with green hatched lines, is always associated with a flash. The convective zone during the helium flash never reaches the outer, hydrogen-rich envelope. The whole phase of subflashes has a duration of about $2$ Myr, in agreement with \citet{Bildsten12}.

The consequence of the helium flash and the subflashes is the presence of small chemical composition gradients in the helium envelope. During each flash a small amount of helium is synthesised into carbon in the $3\alpha$ reaction. The abundance profile of $^{4}\mathrm{He}$ in function of logarithmic relative mass, $\log\,q = \log(m/M)$, in the envelope is shown in Figure\,\ref{fig_he4_envelope}, for the Zero-Age-Extreme-Horizontal-Branch model marked by the vertical line shown in Figure\,\ref{fig_kipp_subflashes}. The abundance $X(^{4}\mathrm{He})$ has a stepped profile in the region where the flashes occurred. The maximum abundance difference between the convective core and the region without prior convection is $\Delta X(^{4}\mathrm{He}) \approx 3.4$. \citet{Constantino15} smoothed out the composition of the envelope before further analysis arguing that the behaviour of helium flashes is dependent on unknown factors related to the extension of convective regions. We do not smooth out the composition in our calculations. The helium flash and subflashes occur in every evolutionary model of post-degenerate subdwarfs and we treat the following abundance pattern as an integral part of these calculations, including the fact that they might have some influence on, e.g., oscillation models.

More details on the helium flash and subflashes in the sdB models calculated with \texttt{MESA} can be found in \citet{Xiong17}.

\subsection{Masses of helium core} \label{sec_progenitors_helium_core}

\begin{figure}
  \includegraphics[clip,width=\columnwidth]{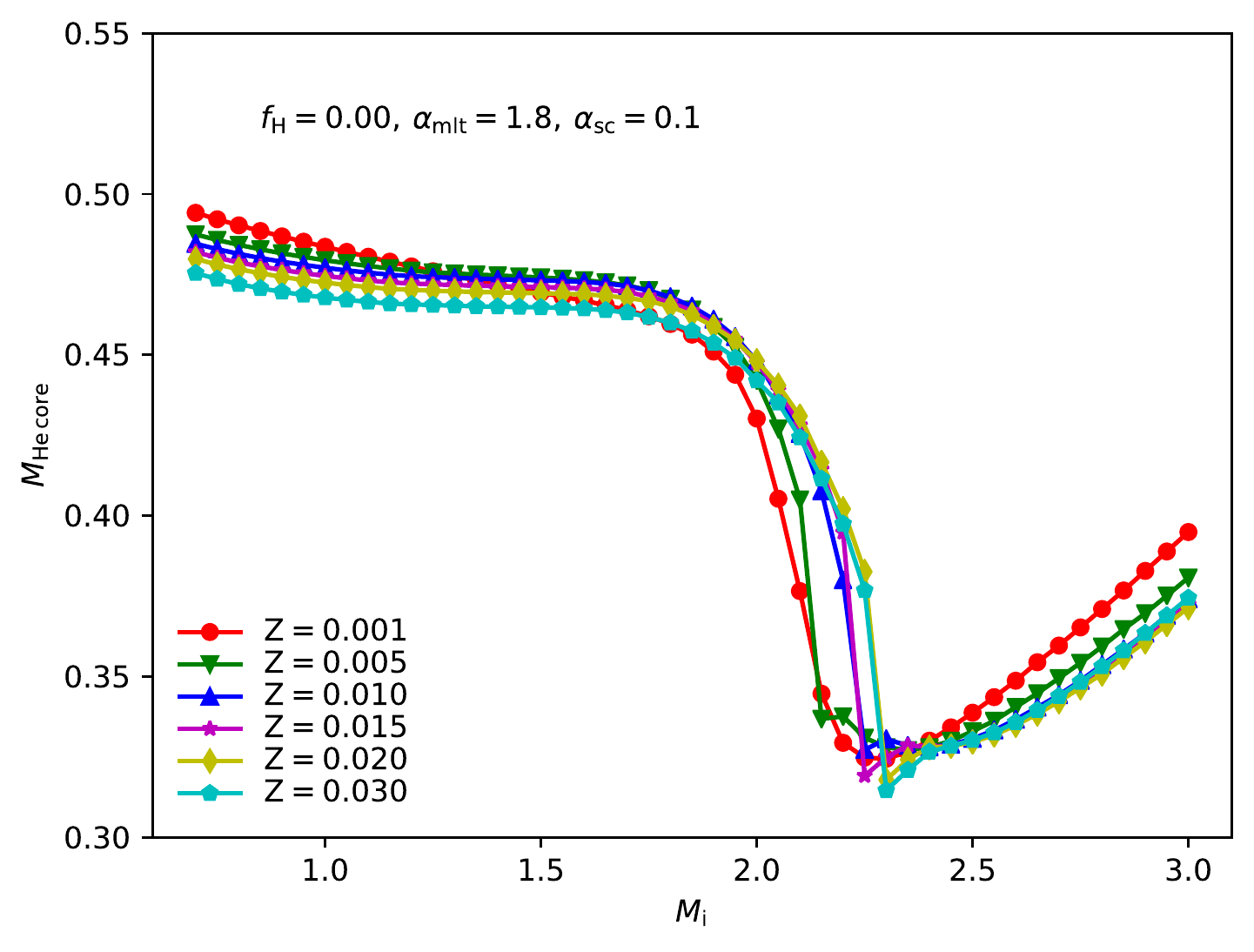}
  \caption{Mass of the helium core at the tip of the red giant branch, $M_\mathrm{He\,core}$, in function of the initial total mass of progenitors, $M_\mathrm{i}$. The effect of metallicity is presented for models with no overshooting from the convective core during the main-sequence evolution. Red dots: models with $Z = 0.001$, green inverted triangles: $Z = 0.005$, blue triangles: $Z = 0.010$, violet stars: $Z = 0.015$, yellow diamonds: $Z = 0.020$, cyan pentagons: $Z = 0.030$.}
  \label{fig_progenitors_z}
\end{figure}

\begin{figure}
  \includegraphics[clip,width=\columnwidth]{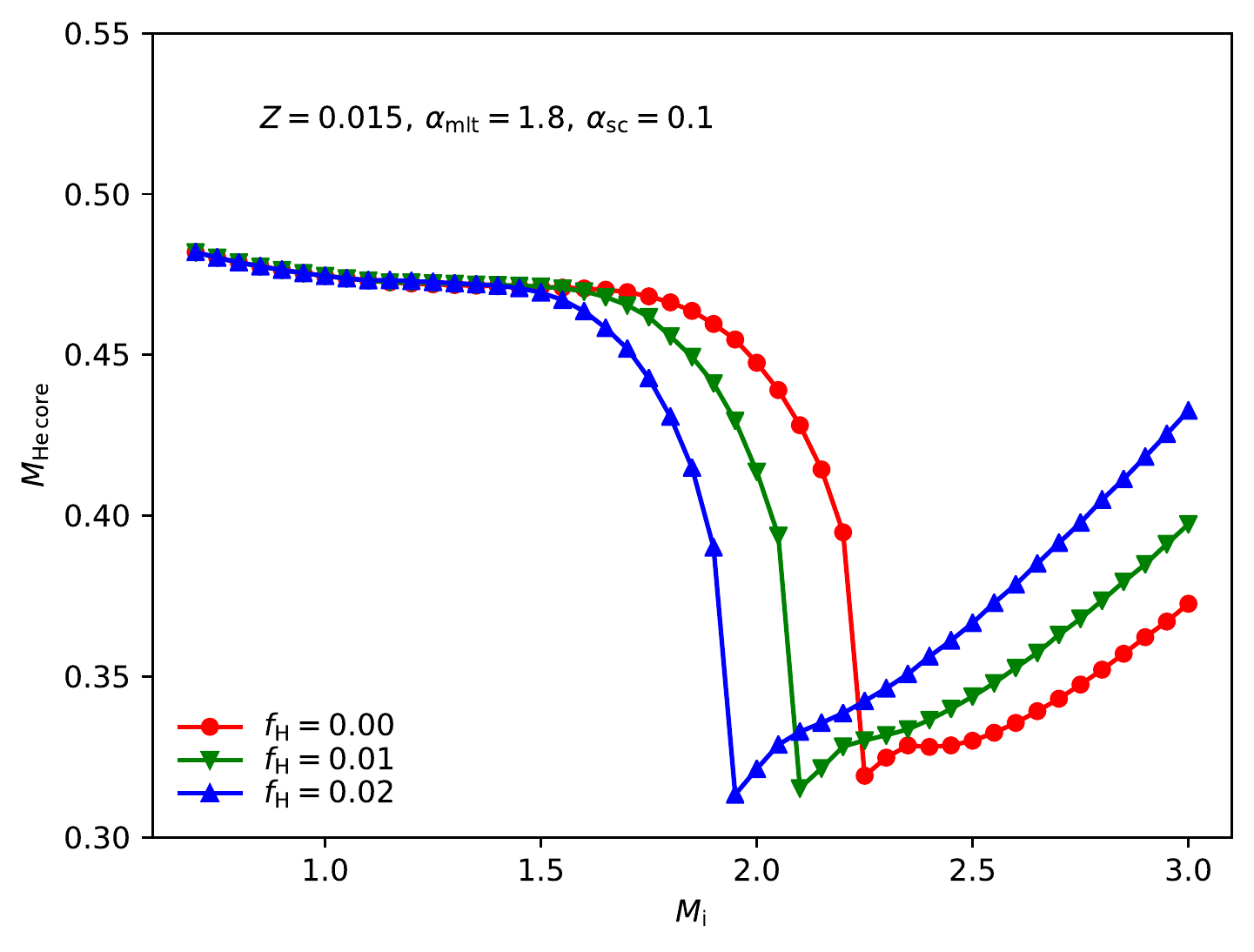}
  \caption{Mass of the helium core at the tip of the red giant branch, $M_\mathrm{He\,core}$, in function of the initial total mass of progenitors, $M_\mathrm{i}$. The effect of convective overshooting from the hydrogen core, $f_\mathrm{H}$, during the main-sequence evolution is shown for models with metallicity $Z = 0.015$. Red dots: $f_\mathrm{H} = 0.0$, green inverted triangles: $f_\mathrm{H} = 0.01$, blue triangles: $f_\mathrm{H} = 0.02$.}
  \label{fig_progenitors_fh}
\end{figure}

The helium-core mass on top of the RGB, $M_\mathrm{He\,core}$, depends on properties of progenitors. This is a very important parameter, because the properties of sdB models depend both on mass of the helium core and the remaining mass of the hydrogen envelope \citep[e.g.][]{Schindler15}. In order to build solid foundations for future work, we have tested how various parameters influence the $M_\mathrm{He\,core}$ for progenitors with initial masses $M_\mathrm{i} = 0.7 - 3.0\,M_\odot$ and  $\Delta M_\mathrm{i} = 0.05\,M_\odot$. All obtained values of $M_\mathrm{He\,core}$ are between $0.31$ and $0.50\,M_\odot$, which is compatible with previous results \citep[e.g.,][]{Han02}.

Metallicity, $Z$, is one of two parameters with the most significant influence on the mass of the helium core. In Figure\,\ref{fig_progenitors_z}, we show the effect of metallicity in a plot with $M_\mathrm{He\,core}$ in function of the initial total mass of progenitors, $M_\mathrm{i}$ for models without overshooting from the hydrogen core. The wide range of metallicities is shown: $Z = 0.001 - 0.030$. The relation between $M_\mathrm{He\,core}$ and $M_\mathrm{i}$ is similar to the results of the previous studies \citep{Sweigart90,Han02,Prada09} and has three distinctive regions, which can be shifted in $M_\mathrm{i}$, especially for different values of $f_\mathrm{H}$. The first is a near-horizontal plateau from the lowest masses up to about $1.9\,M_\odot$ (for the considered parameters). These models have fully degenerated cores during the RGB evolution. Then, there is a steep decrease of $M_\mathrm{He\,core}$, where the core degeneration becomes weaker, until about $M_\mathrm{i} \approx 2.2 - 2.3\,M_\odot$ where the core is no longer degenerated and helium can be ignited without a flash. Finally, there is a linear relation between $M_\mathrm{He\,core}$ and $M_\mathrm{i}$. It is possible to obtain a sdB model with the canonical mass from a non-degenerated progenitor with a mass $M_\mathrm{i} \approx 3.5\,M_\odot$, but it would require extreme mass-loss when compared to the case with degeneration. The differences in $M_\mathrm{He\,core}$ for a given value of $M_\mathrm{i}$ are up to $0.02\,M_\odot$ between $Z = 0.001$ and $0.03$ with the less metallic models achieving higher masses of the helium core. The transition region between degenerated and non-degenerated core is also slightly shifted (up to $0.15\,M_\odot$) towards lower values of $M_\mathrm{i}$ in the models with lower metallicity.

Efficiency of convective overshooting from the hydrogen core, $f_\mathrm{H}$, is the other parameter with a significant impact on $M_\mathrm{He\,core}$. In Figure\,\ref{fig_progenitors_fh}, the relation between $M_\mathrm{He\,core}$ and $M_\mathrm{i}$ is plotted for a fixed metallicity, $Z = 0.015$, and three efficiencies of overshooting: $f_\mathrm{H} = 0.0$, $0.01$ and $0.02$. The considered values of $f_\mathrm{H}$ are in the range expected for the stars with low and intermediate masses \citep[e.g.,][]{Claret17}. Overshooting has no influence on models with masses below $1.1\,M_\odot$, because they do not have convective cores. Between $1.1\,M_\odot$ and $1.5\,M_\odot$, in which the convection starts to emerge in the cores, there is a very small effect on $M_\mathrm{He\,core}$, at the level of about $0.001\,M_\odot$. The effect of overshooting becomes significant in the massive models that do not have degenerated cores on the RGB. For example, for an initial mass of $M_\mathrm{i} = 2.5\,M_\odot$ the obtained helium core masses are $M_\mathrm{He\,core} = 0.323\,M_\odot$ for the case with no overshooting, $0.344\,M_\odot$ for $f_\mathrm{H} = 0.01$, and $0.366\,M_\odot$ for $f_\mathrm{H} = 0.02$. According to the calibration of \citet{Claret17}, the expected value of $f_\mathrm{H}$ for a mass $M_\mathrm{i} = 2.5\,M_\odot$ should be in the range of $0.015$ - $0.02$. It is immediately clear in Figure\,\ref{fig_progenitors_fh} that the transition between models with degenerated and non-degenerated cores during the RGB evolution depends strongly on the main-sequence overshooting and that transition mass is lower for higher values of $f_\mathrm{H}$. In the considered case, the degeneration vanishes at $M_\mathrm{i} = 2.25\,M_\odot$, $2.10\,\,M_\odot$ and $1.95\,M_\odot$ for $f_\mathrm{H} = 0.0$, $0.01$ and $0.02$, respectively.

The mixing-length parameter and the efficiency of semiconvection have very small effect: within the ranges $\alpha_\mathrm{MLT} = 1.6 - 2.2$ and $\alpha_\mathrm{sc} = 0.001 - 1.0$, the differences in $M_\mathrm{He\,core}$ typically do not exceed $2 \times 10^{-3}\,M_\odot$, with the exception of the transition region between degenerate and non-degenerate core, ($M_\mathrm{i} \approx 2.0 - 2.3\,M_\odot$ for $Z = 0.015$ and no overshooting from the hydrogen core). In this mass range differences are higher, up to even $0.025\,M_\odot$ for $M_\mathrm{i} = 2.25\,M_\odot$. This is the first mass with a fully non-degenerated core. Due to almost negligible changes through the considered range of masses we do not present plots for $\alpha_\mathrm{MLT}$ and $\alpha_\mathrm{sc}$.

\end{document}